\documentclass[5p]{elsarticle}
\biboptions{sort&compress}
\usepackage{amsmath}
\usepackage{graphicx}
\usepackage{graphics}
\usepackage{color}
\usepackage{bbold}

\begin{document}

\title{Staggered grid leap-frog scheme for the (2+1)D Dirac equation}
\author[KFU-Graz]{Ren\'{e} Hammer\corref{cor1}}
\ead{rene.hammer@uni-graz.at}
\author[KFU-Graz]{Walter P\"{o}tz}
\ead{walter.poetz@uni-graz.at}
\address[KFU-Graz]{Institut f\"{u}r Physik, Karl-Franzens-Universit\"{a}t Graz, Universit\"{a}tsplatz 5, 8010 Graz, Austria}
\cortext[cor1]{Corresponding author}


\begin{abstract}
A numerical scheme utilizing a grid  which is staggered in both space and time is proposed for the numerical solution of the (2+1)D Dirac equation in presence of an external electromagnetic potential.  It preserves the linear dispersion relation of the free Weyl equation for wave vectors aligned with the grid and facilitates the implementation of open (absorbing) boundary conditions via an imaginary potential term. This explicit scheme has second order accuracy in space and time.  A functional for the norm is derived  and shown to be conserved.  Stability conditions are derived.  Several numerical examples, ranging from generic to specific to textured topological insulator surfaces, demonstrate the properties of the scheme which can handle general electromagnetic potential landscapes.  
\end{abstract}
\begin{keyword}
Dirac equation \sep Topological insulator \sep Finite difference \sep Staggered grid \sep Absorbing boundary conditions
\end{keyword}
\maketitle
\section{Introduction}

Ever since its presentation by P.A.M. Dirac in 1928, the Dirac equation has played a central role in the development of modern physics \cite{dirac}.  It has lead not only to the prediction and observation of antimatter but has also been instrumental to the development of modern many-body physics \cite{dirac, anderson,feynman,itzykson}.  Known as the Weyl equation for zero mass, it has been of relevance to early neutrino physics \cite{ryder}.   While its initial applications naturally were devoted mainly to  high-energy elementary particle physics, it has been known for quite a while that  touching energy bands in crystalline solids also can lead to a Dirac-fermion-like energy dispersion \cite{wallace,kane}.   This can readily be seen for a Schr\"{o}dinger particle in a one-dimensional (1D) periodic potential: Since the spectrum for Bloch solutions has degeneracy two at most, energy bands cannot overlap.  If they touch, they must have linear dispersion near the point of contact \cite{ST-WP}.  
A prominent example for such a situation  is graphene, which has regained great publicity due to its recent experimental realization \cite{novoselov,neto}. For this 2D system, the Brillouin zone features  4 (counting spin) Dirac cones with a small gap due to the spin-orbit interaction.  Odd numbers of Dirac cones have been predicted and experimentally verified on individual  surfaces of topological insulators (TIs) \cite{qi,moore,fu,zhang,xia,hsieh}.  In the simplest case, a single Dirac cone of topologically protected metallic surface states can occur on one side of a TI.   

In a condensed matter environment, effective 2D model systems for our 3D world frequently emerge in the low energy limit. The synthesization of nano-structured materials has led to a number of systems in which electron motion in one spatial dimension is confined  to within a few atomic layers  but  essentially free (quasi-particle) motion over macroscopic length scales occurs in the other two dimensions.    
Celebrated examples  for such genuine (2+1)D systems are the 2DEG and graphene \cite{neto}.  Layered high-T$_c$ systems may be seen as an example in the wider sense \cite{pwanderson}.  Optical lattices also constitute a rich play ground for the engineering of 2D physics with the possibility of tuning various parameters and therewith controlling atom localization and effective many-body interactions \cite{lamata,witthaut,szpak}.
Surfaces of solids in general provide a natural environment for the study of (quasi-) 2D phenomena, with  
a remarkable recent example provided by the  topologically protected 2D metallic surface states of TIs.    Their intrinsically  gap-less energy spectrum can be 
manipulated by perturbations which break time-reversal symmetry to introduce an energy gap (mass term).   Electromagnetic texturing can provide a landscape of electric potential and effective mass, taking positive and negative values,  potentially leading to protected 1D chiral channel states \cite{jackiw,qi,hammerArXiV,hammerAPL}. 

A theoretical analysis of the rich dynamics of  Dirac fermion quasiparticles in (2+1)D requires reliable numerical methods which can handle position- and time- dependent  potential and mass landscapes. Existing methods are the real-space schemes, such as the finite-difference and finite-element methods \cite{becker, stacey, müller}.  Momentum-space spectral methods and split-operator methods have been 
developed also \cite{momberger, braun, mocken, gourdeau}.  While finite-difference and finite-element schemes allow for an easy implementation of non-constant coefficients they have to deal with the fermion doubling problem. This is expressed in the Nielsen-Ninomiya no-go theorem which forbids a single minimum in the energy dispersion of a Dirac-type equation on a regular grid without breaking either Hermiticity, translational invariance, or locality \cite{nielsen}. 
Elimination of fermion doubling by means of  a nonlocal approximation for the spatial derivative operator has been introduced by Stacey \cite{stacey} and implemented numerically for a stationary problem \cite{tworzydlo}.
Fermion doubling can also be avoided by split-operator methods \cite{mocken}.
A scheme with a non-monotonic dispersion relation does not have to be ruled out for numerical studies, but, the latter can severely constrain its useful domain of  wave numbers in momentum space  and may require a very fine grid in real space. Here we will present an easy-to-implement, explicit, finite-difference method which preserves the zero-mass  free Dirac-dispersion of the continuum problem along the main axes of the grid and provides only one extra Dirac cone at the corners of the first Brillouin zone. This  scheme is especially well-suited for long-time propagation studies where the occurring wave vectors mostly are aligned parallel to the grid as envisioned, for example, in Dirac fermion wave guides \cite{hammerAPL}.
Details of the numerical approach and the properties of this scheme are discussed in Sec. \ref{numapp}.  Numerical examples for the free particle propagation, the Klein step, and basic domain wall structures are given in Sect.  \ref{numex}.  Summary and conclusions can be found in Sects. \ref{discu} and \ref{summa}.  Further details are given in the appendix.  

\section{The numerical approach}\label{numapp}
\noindent The generic (2+1)D Dirac equation in normalized units (velocity $c=1$, Planck's constant $\hbar=1$, elementary charge $e = 1$) in Schr\"{o}dinger form may be written as 
\begin{equation}
i \partial_t \mbox{\boldmath$\psi$}(x,y,t) = \hat{H} \mbox{\boldmath$\psi$}(x,y,t)~,
\end{equation}
where $\mbox{\boldmath$\psi$}(x,y,t)\in\mathbb{C}^2$ is a $2$-component spinor and the Hamiltonian is of the form
\begin{equation}
\hat{H} = \sigma_x p_x + \sigma_y p_y  + \mbox{\boldmath$\sigma$}\cdot\mathbf{m}(x,y,t) + V(x,y,t)~,\label{DHAM}
\end{equation}
\begin{figure} [t!]
\centering
\includegraphics[width=8.5cm]{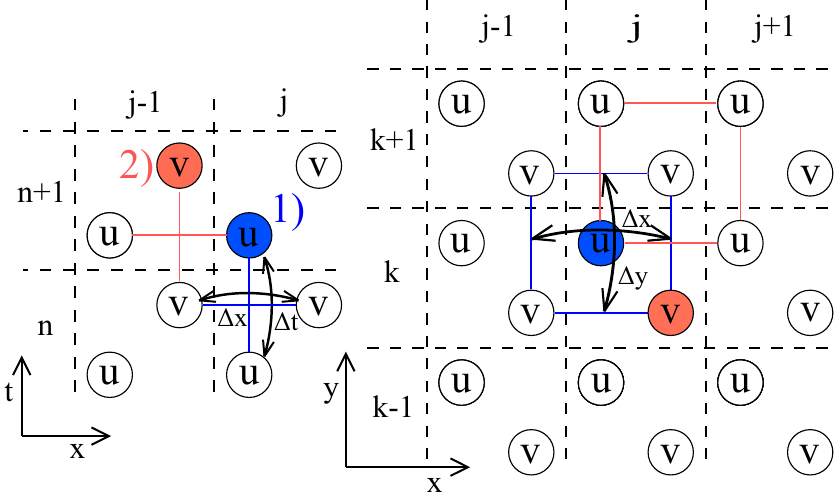}
\caption{(color online). Leap-frog staggered-grid scheme: The left part shows the time-stepping sequence where \textcolor{blue}{1)} the new $u$
components \textcolor{blue}{(blue/dark gray)} are computed by the previous $u$ and the spatial differences of old $v$-values. \textcolor{red}{2)} Then (knowing $u$ at $t_{n+1}$) the new $v$  components \textcolor{red}{(red/light gray)} are computed at $t_{n+1}$ . The right part of this figure shows the pattern for the spatial derivatives in x and y.}
\label{space-time-stepping}
\end{figure}
\noindent $p_i$ represents the component $i=x,y$  of the momentum operator and $\mbox{\boldmath$\sigma$} = (\sigma_x,\sigma_y,\sigma_z)$ is the vector of Pauli matrices. The scalar potential in Eq. \eqref{DHAM} is represented by $V$. The ``magnetization vector''\\ $\mathbf{m}(x,y,t)$ $=[m_x(x,y,t),m_y(x,y,t),m_z(x,y,t)]$ may have its origin in a vector potential or an exchange coupling to a ferromagnetic medium.   Note that $m_z\neq 0$ provides the mass term to the equation.
  
This generic two-component  Dirac equation provides an effective model for a number of physical systems.  
 For topological insulators, this model describes the low energy surface excitations.  The spin is locked to the momentum, whereby the physical spin quantization axis is $\mathbf{S}\propto\mathbf{\hat z}\times \mbox{\boldmath$\sigma$}$, corresponding to a fermion with 2 degrees of freedom \cite{qi}.
For given momentum,  the presence of two components may be interpreted as accounting either for the existence of positive and negative energy solutions or the presence of two spin directions.   Note, that flipping the spin for given momentum is equivalent to switching between the positive to negative energy branch.
 Indeed, by a simple unitary transformation applied to Eq. \eqref{DHAM} one can arrive at a ``physical" representation such that  $\mathbf{S}\propto \mbox{\boldmath$\sigma$}$ which is useful in the presence of external electromagnetic fields.  Note that,  in accordance with the Nielsen-Ninomiya no-go theorem,  a second Dirac cone is located on the opposite side of the TI surface.  
For graphene, the two components arise from the $|p_z\rangle$ bonding and anti-bonding band (without spin) \cite{neto}.   
In contrast, the standard four-component Dirac equation describes a spin-1/2 particle (Dirac fermion) with positive and negative energy solutions amounting to $2\times 2=4$ degrees of freedom (``two energy bands and two spin directions"). The Majorana fermion solution is a special case.  It is its own antiparticle, making it ``half a Dirac fermion" with two degrees of freedom \cite{sakurai,greiner,ryder,itzykson,srednicki}.  
\\
\subsection{Numerical scheme}

\noindent We propose a staggering of the grid both in time and space with second-order  approximation for the time and space derivative, as shown in Fig. \ref{space-time-stepping}, %
\begin{align}
&\frac{u_{j,k}^{n+1}-u_{j,k}^{n}}{\Delta t}= -i\left((m_z)^n_{j,k}+V^n_{j,k}\right)\frac{u_{j,k}^{n+1}+u_{j,k}^{n}}{2} \nonumber \\\nonumber
&-\frac{(v_{j,k-1}^{n}-v_{j-1,k-1}^{n})+(v_{j,k}^{n}-v_{j-1,k}^{n})}{2 \Delta x}\\\nonumber
&+i\frac{(v_{j-1,k}^{n}-v_{j-1,k-1}^{n})+(v_{j,k}^{n}-v_{j,k-1}^{n})}{2 \Delta y}~,\\\nonumber
\\\nonumber
&\frac{v_{j,k}^{n+1}-v_{j,k}^{n}}{\Delta t}=i\left((m_z)^{n+1}_{j,k}-V^{n+1}_{j,k}\right)\frac{v_{j,k}^{n+1}+v_{j,k}^{n}}{2}\\\nonumber
&-\frac{(u_{j+1,k}^{n+1}-u_{j,k}^{n+1})+(u_{j+1,k+1}^{n+1}-u_{j,k+1}^{n+1})}{2 \Delta x}\\
&-i\frac{(u_{j,k+1}^{n+1}-u_{j,k}^{n+1})+(u_{j+1,k+1}^{n+1}-u_{j+1,k}^{n+1})}{2 \Delta y}~\label{lfsgscheme}
\end{align}
with the notation $\mbox{\boldmath$\psi$}(x_j,y_k,t_n) \approx \mbox{\boldmath$\psi$}_{j,k}^n = (u_{j,k}^n,v_{j,k}^n)$ where $n$ and $j,k$, respectively,  are the discrete time and space indices.

Here the mass term $m_z$ and the potential $V$ enter the scheme in a Crank-Nicolson time-averaged manner.  A consistent incorporation of a  "vector potential" $m_x$ and $m_y$ will be detailed below. Since the time averaged functions for the former only depend on one spatial grid-point a rearranging of terms leads to an explicit scheme. We call it a leap-frog scheme because $u$ and $v$ are computed in an alternating manner where first $u_{j,k}^{n+1}$ is computed from  $u_{j,k}^{n}$ and $v_{j,k}^{n}$. Then using the updated components $u_{j,k}^{n+1}$, as shown in Fig. \ref{space-time-stepping},  the new $v_{j,k}^{n+1}$ are computed. This spatial staggering allows for a centered approximation of the first spatial partial derivatives without omitting the central grid point,  as is the case for a centered symmetric first derivative operator on a regular grid. This eliminates one source of fermion-doubling. Here it should be recalled that by using  one-sided difference operators with alternating direction for for u and v, fermion doubling can be avoided for the (1+1)D Dirac equation.  For the (1+1)D case the latter is equivalent to the present spatial staggering of the grid \cite{stacey, hammer1}. We use staggering in time to further improve the dispersion relation, which will be shown below.\\
\\
$m_x$ and $m_y$ terms in the Hamiltonian Eq. \eqref{DHAM} are incorporated consistently into the scheme above, for $m_x=m_y=0$, using a Peierls substitution\cite{peierls,graf},
\begin{eqnarray}
u^{n}_{j,k} \rightarrow  {\hat u}^{n}_{j,k} \equiv u^{n}_{j,k} \exp\{i   a^{n}_{j,k}\} \nonumber \\
v^{n}_{j,k} \rightarrow {\hat v}^{n}_{j,k} \equiv  v^{n}_{j,k} \exp\{i  a^{n}_{j,k} \}~,
\label{peierlss}
\end{eqnarray}
where the real phase $a^{n}_{j,k}$ is defined as the line integral over the two-dimensional magnetization vector ${\bf m}$, starting at arbitrary, but fixed position $(x_o,y_o)$ and ending on the lattice point 
$(x,y)=(j\Delta x, k\Delta y)$,
$$
a^{n}_{j,k} =  \int_{(x_o,y_o)}^{(x,y)} d{\bf s}\cdot {\bf m}({\bf s},t)\mid_{x=j\Delta x, y=k\Delta y, t=n\Delta t}~.
$$
This substitution introduces $m_x$ and $m_y$ in covariant fashion when interpreted as components of the electromagnetic vector potential, leading to $p_x\rightarrow p_x+m_x$ and $p_y\rightarrow p_y+m_y$ in the Dirac Hamiltonian Eq. \eqref{DHAM}.  Details of the resulting scheme are discussed in \ref{A}.  In the limit of smooth variation of $m_x$ and $m_y$ in space and time the scheme takes the form
\begin{align}
&\frac{u_{j,k}^{n+1}-u_{j,k}^{n}}{\Delta t}=- i\left((m_z)^{n}_{j,k}+\hat{V}^{n}_{j,k} \right)\frac{u_{j,k}^{n+1}+u_{j,k}^{n}}{2}  \nonumber \\
& -\frac{v_{j,k-1}^{n}-v_{j-1,k-1}^{n}+v_{j,k}^{n}-v_{j-1,k}^{n}}{2 \Delta x} \nonumber \\
 &+ i \frac{v_{j-1,k}^{n}-v_{j-1,k-1}^{n}+v_{j,k}^{n}-v_{j,k-1}^{n}}{2\Delta y} \nonumber \\
 &- i(m_x)^{n}_{j,k}  \frac{v_{j,k}^{n}+v_{j-1,k}^{n}+ v_{j,k-1}^{n}+v_{j-1,k-1}^{n}}{4}  \nonumber \\
 &-(m_y)^{n}_{j,k} \frac{v_{j-1,k}^{n}+v_{j-1,k-1}^{n}+v_{j,k}^{n}+v_{j,k-1}^{n}}{4}~, \label{lfsgscheme-u}
\end{align}
and
\begin{align}
&\frac{v_{j,k}^{n+1}-v_{j,k}^{n}}{\Delta t}= i\left((m_z)_{j,k}^{n+1}-\hat{V}_{j,k}^{n+1} \right)\frac{v_{j,k}^{n+1}+v_{j,k}^{n}}{2}  \nonumber \\
&- \frac{u_{j+1,k}^{n+1}-u_{j,k}^{n+1}+u_{j+1,k+1}^{n+1}-u_{j,k+1}^{n+1}}{2\Delta x} \nonumber \\
&-i\frac{u_{j,k+1}^{n+1}-u_{j,k}^{n+1}+u_{j+1,k+1}^{n+1}-u_{j+1,k}^{n+1}}{2\Delta y} \nonumber \\
& -i(m_x)^{n+1}_{j,k}\frac{u_{j+1,k}^{n+1}+u_{j,k}^{n+1}+u_{j+1,k+1}^{n+1}+u_{j,k+1}^{n+1}}{4} 
\nonumber  \\
 &+ (m_y)^{n+1}_{j,k}\frac{u_{j,k+1}^{n+1}+u_{j,k}^{n+1}+u_{j+1,k+1}^{n+1}+u_{j+1,k}^{n+1}}{4} ~.\label{lfsgscheme-v}
\end{align}
Here $\hat{V}_{j,k}^{n+1} $ is the net scalar potential.  Its relation to ${V}_{j,k}^{n+1} $ is is given in \ref{A}.

\subsection{Von Neumann stability analysis}
\begin{figure}[t!]
\centering
\includegraphics[width=8.5cm]{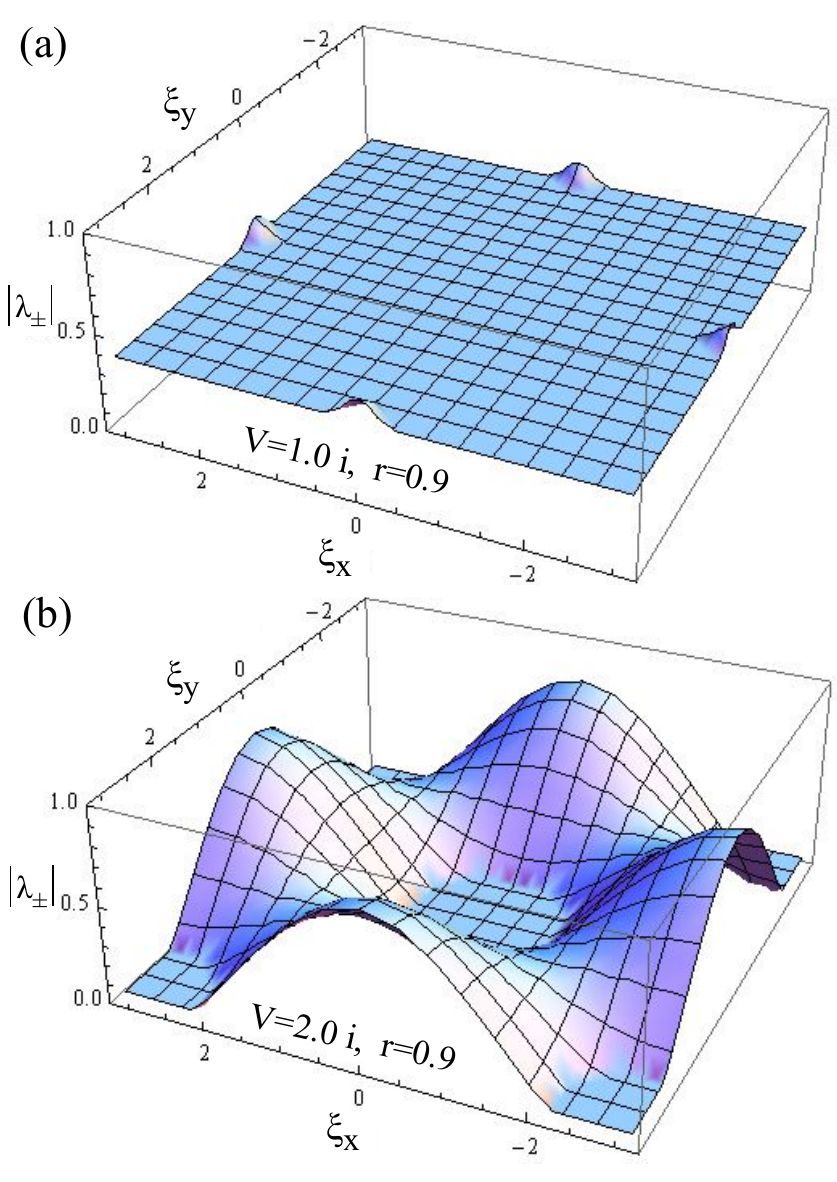}
\caption{(color online). The largest eigenvalue of the growth matrix $G$ max$(|\lambda_\pm|)$ shown over
the entire scaled $k$-space 
for $m=0$ and two different values of imaginary $V$, for ratio $r=\Delta t/\Delta=0.9$.}
\label{lambda-imag-V}
\end{figure}
For this linear system and constant coefficients,  Fourier analysis is used to determine the dispersion introduced by the grid.  Furthermore,  we use periodic boundary conditions (absorbing layers as introduced  later do not violate periodicity).  Thus, von Neumann stability analysis is sufficient to explore the stability of the finite difference scheme \cite{strikwerda}.  
The Fourier transform of Eq. (3) from real space to momentum space leads to:
\begin{equation}
\underbrace{\left(\begin{array}{cc}
a_{11} & a_{12}\\
a_{21} & a_{22} 
\end{array}\right)}_{=:A}
\left(\begin{array}{c}
\tilde{u}^{n+1}\\
\tilde{v}^{n+1}
\end{array}\right)
+\underbrace{\left(\begin{array}{cc}
b_{11} & b_{12} \\
b_{21} & b_{22}  
\end{array}\right)}_{=:B}
\left(\begin{array}{c}
\tilde{u}^{n}\\
\tilde{v}^{n}
\end{array}\right)=0~\label{stabilityeq},
\end{equation}
where we define 
\begin{align}
a_{11} =&\frac{1}{\Delta t}+\frac{i(m_z+V)}{2}~,\\\nonumber
a_{12} =& 0~,\\\nonumber
a_{21} = &+\frac{\left(1-e^{i k_x \Delta_x}\right) \left(1+e^{i k_y \Delta_y}\right)}{2 \Delta x}\\\nonumber
&-i\frac{\left(1+e^{i k_x \Delta_x}\right) \left(1-e^{i k_y \Delta_y}\right)}{2 \Delta y}\\\nonumber
&+i \frac{(m_x-i m_y)\left(2+e^{i k_x \Delta_x}+e^{i k_y \Delta_y}\right)}{4}~,\\\nonumber
a_{22} =& \frac{1}{\Delta t}-\frac{i(m_z-V)}{2}~,\\\nonumber
b_{11} =& -\frac{1}{\Delta t}+\frac{i(m_z+V)}{2}~,\\\nonumber
b_{12} =& -\frac{\left(1-e^{-i k_x \Delta_x}\right) \left(1+e^{-i k_y \Delta_y}\right)}{2 \Delta x}\\\nonumber
&-i\frac{\left(1+e^{-i k_x \Delta_x}\right) \left(1-e^{-i k_y \Delta_y}\right)}{2 \Delta y}\\\nonumber
&+i \frac{(m_x+i m_y)\left(2+e^{-i k_x \Delta_x}+e^{-i k_y \Delta_y}\right)}{4}~,\\\nonumber
b_{21} =& 0~,\\\nonumber
b_{22}=&-\frac{1}{\Delta t}-\frac{i(m_z-V)}{2}~.
\end{align}\\
\noindent It is convenient to define the amplification matrix $G = -A^{-1} B~$ and to use $\xi_x = k_x \Delta_x$, $\xi_y = k_y \Delta_y$ Eq. \eqref{stabilityeq} becomes
\begin{equation}
\tilde{\psi}^{n+1}(\xi_x,\xi_y)=G(\xi_x,\xi_y) \tilde{\psi}^{n}(\xi_x,\xi_y)~.\label{Gshort}
\end{equation}
We now consider $\Delta x = \Delta y = \Delta$ and define the ratio $r=\Delta t/\Delta$. Introducing rescaled variables $\mu_i = m_i \Delta t$ and $\nu = V \Delta t$, we write the eigenvalues of $G$ using the root formula
\begin{equation}
\lambda_\pm=  P/2 \pm \sqrt{\big(P/2\big)^2-Q}~,
\label{growthfactor}
\end{equation}\\
where $P = \mbox{tr} [G]$ and $Q = \mbox{det}[G]$.
The resulting lengthy expressions write as:
\begin{align}
P=&\Big[16 (1-r^2) + 4(\nu^2-\mu_z^2) - 3 (\mu_x^2+\mu_y^2)\\\nonumber
&+4 r (\mu_x-\mu_y)  - 2 (\mu_x^2+\mu_y^2)(\cos \xi_x + \cos \xi_y)\\\nonumber
&+4 r (\mu_x+\mu_y)(\cos \xi_x - \cos \xi_y)\\\nonumber
&-(\mu_x^2+\mu_y^2)(\cos \xi_x \cos \xi_y + \sin \xi_x \sin \xi_y)\\\nonumber
&-4 r(\mu_x-\mu_y)(\cos \xi_x \cos \xi_y - \sin \xi_x \sin \xi_y )\\\nonumber
&+8 r\big(\mu_x \sin \xi_x + \mu_y \sin \xi_y + \mu_x \sin \xi_x \cos \xi_y\\\nonumber
&+ \mu_y \cos \xi_x \sin \xi_y + 2 r \cos \xi_x \cos \xi_y\big)\Big]/(2 N)~,\\\nonumber
Q=&[\mu_z^2-(\nu-2 i)^2]/N\label{Q}~,
\end{align}
\noindent where $N=4+\mu_z^2-4 i \nu-\nu^2$.
For $\mu_x=\mu_y=0$ it reduces to: $P=[2(\nu^2-\mu_z^2)+8 r^2 (1/r^2-1+\cos \xi_x \cos \xi_y)]/N$.
Now one can show that for $r<1$ and $\mu_z,\nu \in\mathbb{R}$: $\left|\lambda_\pm\right|=1$ for all allowed wave numbers on the grid, $\xi_x, \xi_y \in \left[-\pi,\pi\right]$. Since under this constraints the eigenvalues are simple the the scheme is stable for the ratio $r < 1$, which constitutes the CFL condition for the scheme. For $\mu_i\neq0$ the scheme is stable for $r<\mbox{min}\big[\sqrt{1-(\mu_x/4)^2}-\mu_y/4,\sqrt{1-(\mu_y/4)^2}+\mu_x/4\big]$. Let's, for now and in momentum space,  define the square of the $l_2$ norm of the spinor as:\\ $\|\mbox{\boldmath$\tilde{\psi}$}\|^2$ $= \int_{-\pi}^\pi \mathrm{d}\xi_x \int_{-\pi}^\pi  \mathrm{d}\xi_y (|\tilde{u}(\xi_x,\xi_y)|^2+|\tilde{v}(\xi_x,\xi_y)|^2)$.  It is preserved because $\|G \mbox{\boldmath$\tilde{\psi}$}\|=\|\mbox{\boldmath$\tilde{\psi}$}\|$. If $m_z$ has an imaginary part, it turns out that the scheme is unstable because the absolute value of one of the $\lambda$'s becomes larger than one in the vicinity of the points $(\xi_x,\xi_y)=(0,0)$ and $(\xi_x,\xi_y)=(\pm\pi,\pm\pi)$.  If, on the other hand,  the potential $V$ has an imaginary part its sign determines the stability of the scheme:  for  $\Im\{V\}<0$ it is unstable.  
For $\Im\{V\} >0$ it is stable, however, $\mbox{max}(\left|\lambda_\pm\right|)<1$ and the norm is not conserved.
One can utilize this property to create absorbing layers which act as absorbing (open) boundary conditions within this scheme. Furthermore, internal absorbing layers can be used to simulate particle source and drain. In Fig. \ref{lambda-imag-V} we show, for illustration,  max$(|\lambda_\pm|)$ and its behavior for different values of $r$ and $\Im\{V\}>0$.\\
\\
\subsection{Dispersion relation}
\begin{figure}[t!]
\centering
\includegraphics[width=8.5cm]{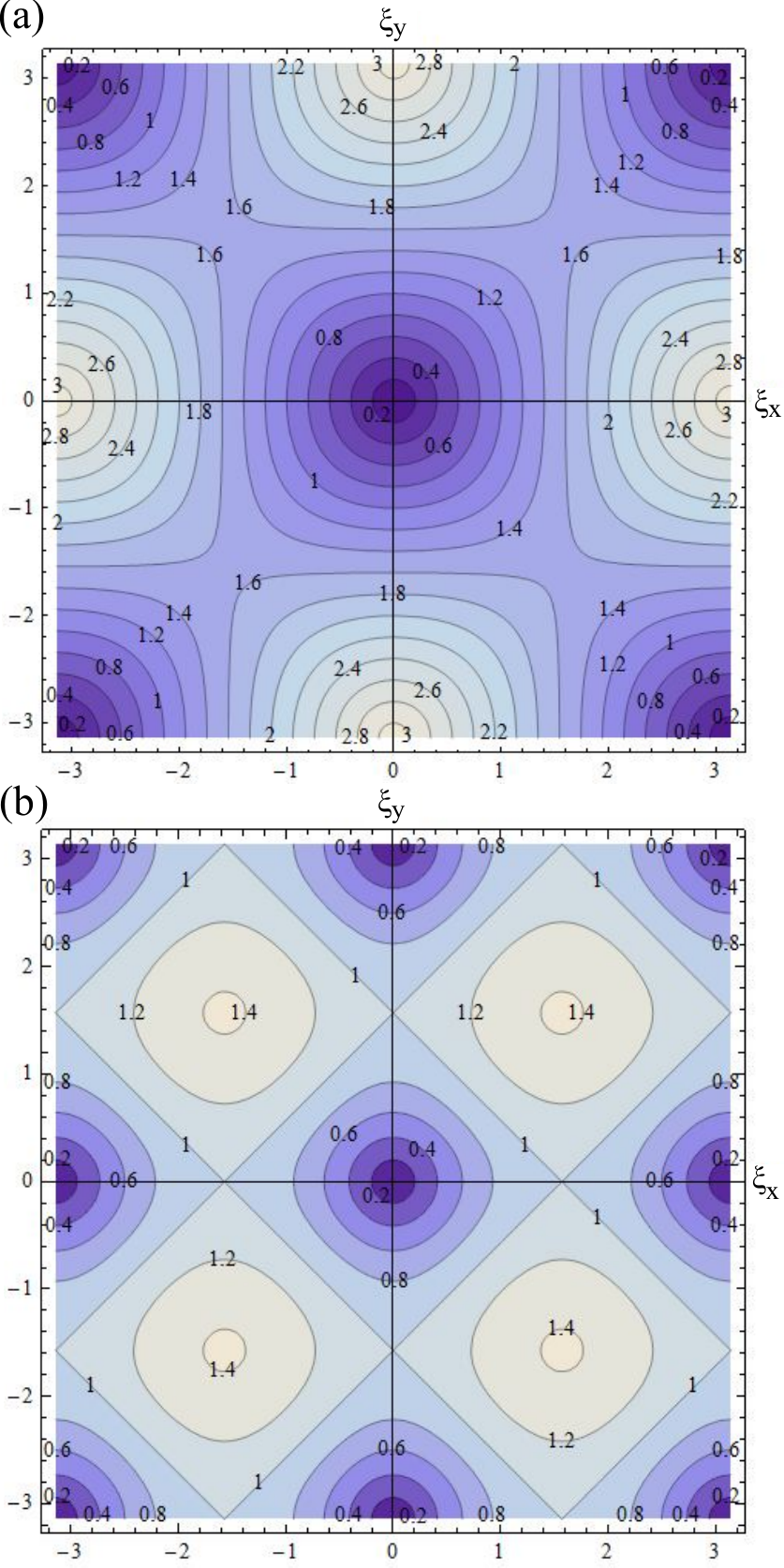}
\caption{(color online). Dispersion relation for $\Delta x = \Delta y = \Delta$, $\mu_i = 0$, $\nu = 0$ and $r=1$. (a) Leap-frog staggered-in-space-and-time scheme. For comparison: (b) centered differences in space and Crank-Nicolson in time, without grid staggering.}
\label{disp-lf}
\end{figure}
\noindent Let us now turn our attention to the dispersion relation of the proposed scheme.
It is obtained by a Fourier transformation  of Eq. \eqref{Gshort} in time or,  again by looking at the eigenmodes of the scheme. The ansatz: $\tilde{u}^{n+1} = e^{i \tilde{\omega}} \tilde{u}^{n}$ where $\tilde{\omega}=\omega \Delta t$,  gives us, in the notation from above, the homogeneous system
\begin{equation}
(e^{i \tilde{\omega}} A + B) \mbox{\boldmath$\tilde{\psi}$} = 0~,
\end{equation}
from which the non-trivial solutions are determined for $\tilde{\omega}$ and expressed in terms of  the growth factor $\lambda_{\pm}$ (see Eq. \eqref{growthfactor})
\begin{equation}
\tilde{\omega} = -\frac{i}{r} \ln [\lambda_{\pm}]~\label{grid-dis}.
\end{equation}  
Here again the necessary and sufficient condition for the conservation of the norm, using  $\lambda_{\pm}^n=e^{i \omega r n}$, shows up as $\Im (\omega) = 0$ which, for $\mu_x,\mu_y=0$, gives  $|\lambda_{\pm}|=1$. 
Setting $\mu_z=0$ and $\nu=0$ it simplifies to
\begin{align}
\tilde{\omega} = - \frac{i}{r} \ln \Bigg\{&1 + r^2 \big(\cos\xi_x \cos\xi_y-1\big)\\\nonumber 
&\pm \sqrt{\Big[ r^2 \big(\cos\xi_x \cos\xi_y-1\big)+1\Big]^2-1} \Bigg\},
\label{grid-dis}
\end{align}  
\noindent where choosing $r=1$ leads to:
\begin{equation}
\tilde{\omega} = - i \ln \Big(\cos\xi_x \cos\xi_y \pm \sqrt{\cos^2\xi_x \cos^2\xi_y-1} \Big)~,
\end{equation}  
which is plotted in Fig. \ref{disp-lf}.
It can be seen that the dispersion relation is linear along the $x$- and $y$-axis. For example, at the $x$ axis where $\xi_y=0$,  we get
\[
\tilde{\omega} = -i \ln \Big(\cos \xi_x \pm i \sin \xi_x \Big) = \pm \xi_x ~.
\]

\begin{figure*}[t!]
\centering
\includegraphics[width=18cm]{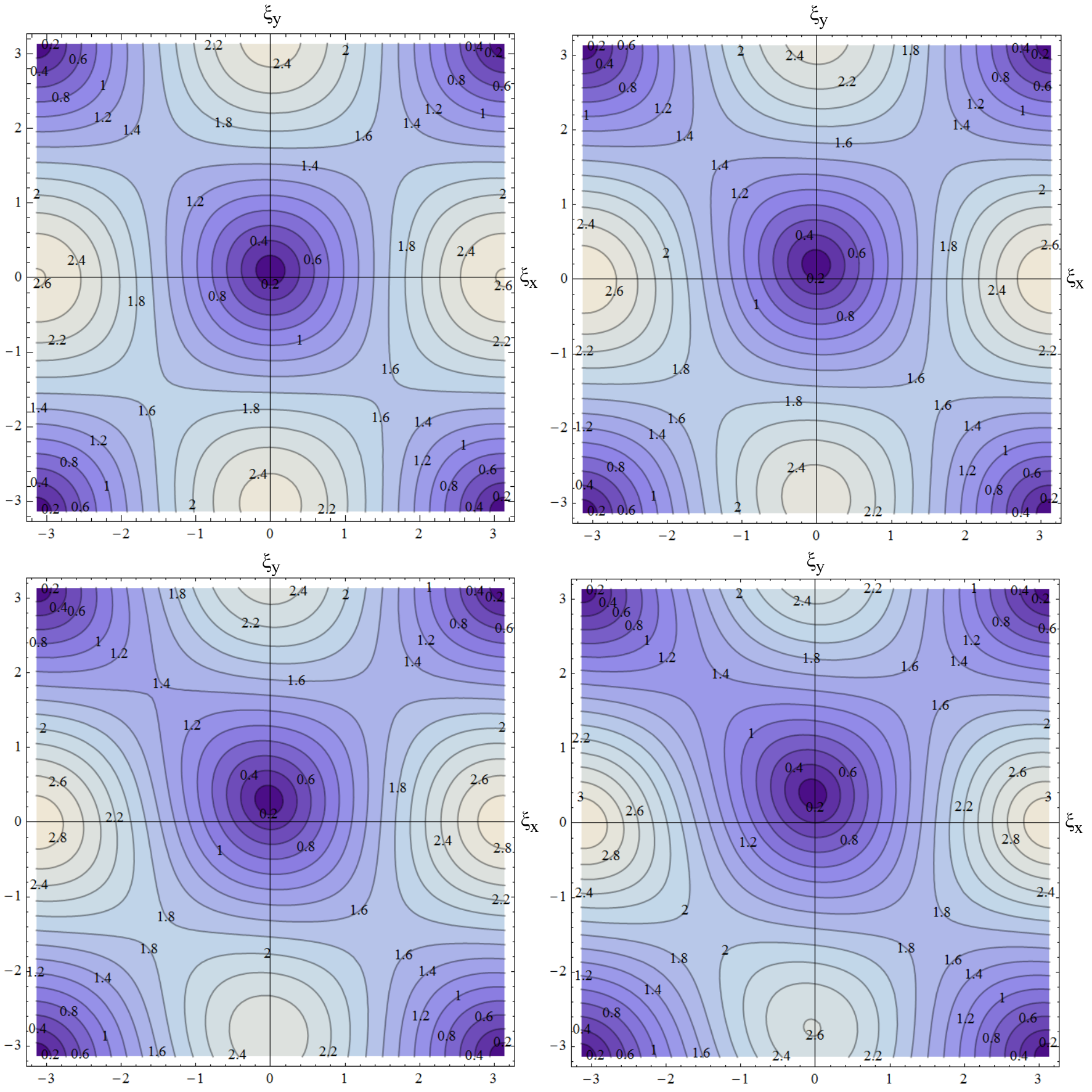}
\caption{(color online). Dispersion relation of the present leap-frog scheme with $r=0.9$, $\Delta x = \Delta y = \Delta$, $m_x=m_z=V= 0$ and $m_y = 0.1, 0.2, 0.3, 0.4$ from the upper left to the lower right figure.  One observes a shift of the Dirac cone away from the center of the momentum space by $m_y$ consistent with the continuum solution.}
\label{disp-with-my}
\end{figure*}

\noindent This means that the dispersion relation of the continuum model (Weyl equation) is preserved along the $x$- and $y$- axis. For comparison we also show the dispersion relation for a scheme using a centered second-order approximation without spatial staggering of the spatial differences and without time staggering but using Crank-Nicolson in time instead (Fig. \ref{disp-lf} (b)). With the staggered scheme one gets two Dirac cones, at $(\xi_x,\xi_y)=(0,0)$ and $(\xi_x,\xi_y)=(\pm\pi,\pm\pi)$ respectively, in contrast to the four Dirac cones obtained by the scheme without staggering. 
The dispersion relation for $\mu_y\neq 0$ is computed numerically and shown in Fig. \ref{disp-with-my}.\\
\\

\begin{figure*}[t!]
\centering
\includegraphics[width=13cm]{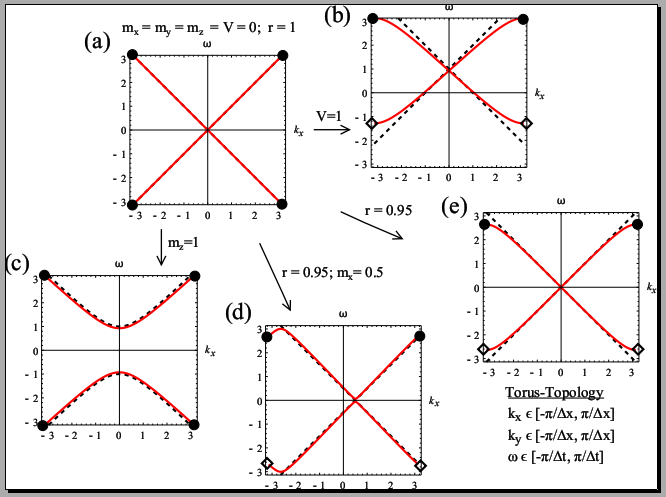}
\caption{(color online). Dispersion relation for wave vectors aligned with the $k_x$ axis ($k_y=0$), $\Delta x =1$ for various parameters. The topology of the dispersion relation is that of a torus, periodic in $k_x$,  $k_y$, and $\omega$.  In each figure, equivalent points in the dispersion are marked by pairs of black filled circles and empty diamonds, respectively.   For comparison, the dispersion relation for the continuum equation is shown by dashed lines.}
\label{disp-topology}
\end{figure*}

Putting a genuine continuum model on a lattice inevitably leads to changes in the spectral properties.  The spatial grid destroys momentum conservation of the free-particle Dirac equation and introduces an invariance under discrete (primitive) translations in real space.  As a consequence, $k$-vectors are defined up to reciprocal lattice vectors only.  Similarly, 
a time grid with spacing  $\Delta t$ makes frequency well defined only within the interval $(-\pi/\Delta t, +\pi/\Delta t]$. 
One remarkable feature of the present model is that for the special case of $r=1$ the linear energy dispersion of the free-particle Weyl equation is preserved exactly along $k_x$ and $k_y$.  This is shown in Fig.  \ref{disp-topology} (a).  No phase errors occur in this case, which, has recently been used to study the dynamics of Dirac fermions in a 2D interferometer setup \cite{hammerAPL}.  
As expected, however, errors (deviation from continuum behavior - dashed lines in Fig. \ref{disp-topology}) occurs in all other cases: 
 Adding a (constant) potential $V$, 
 Fig. \ref{disp-topology} (b), introduces an artificial ``energy gap" at the BZ boundary while the correct dispersion should remain linear, merely shifted vertically by $V$.    In this case, periodic motion (Bloch oscillation) occurs when the k-vector crosses the BZ boundary.  
Similarly for finite mass,  Fig. \ref{disp-topology} (c), a wave packet will disperse while undergoing Bloch oscillations.  
Errors in the dispersion also occur for $r<1$,  Fig. \ref{disp-topology} (e).  The effect of $m_x\neq 0$ is shown in Fig. Fig. \ref{disp-topology} (d).  While it changes the kinetic momentum leading to a horizontal shift of the dispersion in the continuum model, additional band curvature arises on the lattice.  

\subsection{The norm}
\begin{figure}[t!]
\centering
\includegraphics[width=8.5cm]{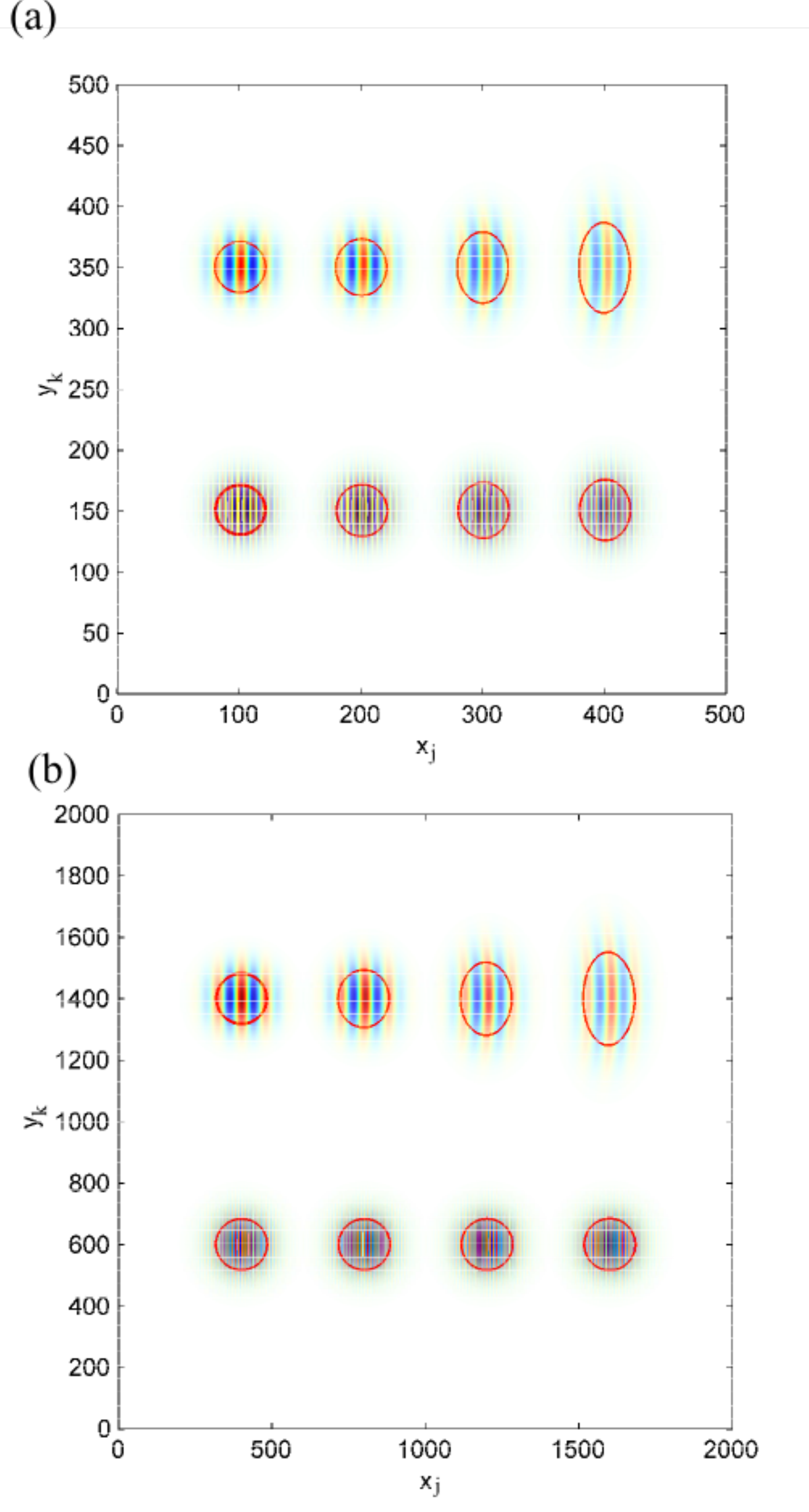}
\caption{(color online). (a) Comparison of the propagation of a packet for $m=V=0$ with a low wave numbers $(k_x,k_y)=(10.0\pm1.3,0.0\pm1.3)\%$ to one prepared with a mean wave number near the maximum for the grid $(k_x,k_y)=(80.0\pm1.3,0.0\pm1.3)\%$. (b) The same initial data using a finer grid having $(k_x,k_y)=(2.50\pm0.32,0.0\pm0.32)\%$ and $(k_x,k_y)=(20.0\pm0.32,0.0\pm0.32)\%$.  The closed lines represent the FWHM the brightness saturation the probability density and the color/brightness variation encodes the phase.}\label{sim_racecoarsefine}
\end{figure}
\noindent The definition of a discrete $L_2$ norm on a staggered grid requires some care due to ambiguities when taking the continuum limit. In particular, the simple local  form
\begin{equation}
||\psi||_2^n:= \sqrt{\sum_{j,k} \left(|u_{j,k}^n|^2 + |v_{j,k}^n|^2\right)}\label{badnorm}
\end{equation}
proves to be a poor choice since numerical tests show that, while conserved on time average,  it can show strong oscillations around its mean value.

In order to define a norm which is invariant under this scheme we first define a scalar product between spinor components on the lattice as follows

\begin{equation}
( u^n ; v^{n'})_{0,0}=( u^n ; v^{n'})= \sum_{j,k} (u_{j,k}^n)^* v_{j,k}^{n'}~ ,\label{scp1}
\end{equation}
and
\begin{equation}
|| u^n||^2= ( u^n ; u^{n})~.
\end{equation}
Furthermore we define scalar products with shifted spinor components as follows
\begin{align*}
( u^n ; v^{n'})_{\pm,0}&=( u^n ; v^{n'}_{\pm,0}) = \sum_{j,k} (u_{j,k}^n)^* v_{j\pm1,k}^{n'}~,\\
( u^n ; v^{n'})_{0,\pm}&= ( u^n ; v^{n'}_{0,\pm}) = \sum_{j,k} (u_{j,k}^n)^* v_{j,k\pm1}^{n'}~,\label{scp2}
\end{align*}
and
\begin{equation}
( u^n ; v^{n'})_{\pm,\pm}= ( u^n ; v^{n'}_{\pm,\pm}) = \sum_{j,k} (u_{j,k}^n)^* v_{j\pm1,k\pm1}^{n'}~,\label{scp3}
\end{equation}

Here the sum $j,k$ runs over all lattice sites.  $n$, $n'$ denote two time sheets, and $u$  and $v$ denote any combination of two upper and/or lower spinor components.

For any physical situation, we may consider either zero boundary conditions or periodic boundary conditions.  In both cases, a norm which  is conserved by the scheme Eqs. \eqref{lfsgscheme} for real-valued and finite $m_z(x,y,t)$ and $V(x,y,t)$, and $m_x=m_y=0$  is given by
\begin{align}
E_{n+1}=E_n=&|| u^n||^2 + ||v^n||^2 \nonumber \\
&- r_x\Re\big\{ (u^n;v^n)_{0,-} - (u^n;v^n)_{-,-}\nonumber  \\
&+ (u^n;v^n)_{0,0} - (u^n;v^n)_{-,0} \big\} \nonumber  \\
&- r_y\Im\big\{ (u^n;v^n)_{-,0} - (u^n;v^n)_{-,-}\nonumber  \\
&+ (u^n;v^n)_{0,0}-(u^n;v^n)_{0,-} \big\} ~.\label{norm}
\end{align}
with $r_x=\Delta t/(2 \Delta x)$ and $r_y=\Delta t/(2 \Delta y)$. 

Furthermore, for the general case of non-vanishing  $m_x$ and $m_y$, the conserved norm $\hat{E}_n$ under the exact scheme (see \ref{A} and \ref{B}) is obtained by subjecting $u$ and $v$ in $E_n$ to the Peierls transformation  Eq. \eqref{peierlss}, noting that $\exp\{i  a^{n}_{j,k}\}$ is a local gauge field.  Note also that the scheme Eq. \eqref{lfsgscheme-u} and Eq. \eqref{lfsgscheme-v} is valid 
only in the limit of slowly-varying $m_x$ and $m_y$ and hence will conserve $\hat{E}_n$ only in this limit.  The proof for the conservation of $E_n$ and, respectively, $\hat{E}_n$ is given in \ref{B}.  
\\

This definition of a norm also allows a stability analysis of the scheme.  In particular, it is valid for time- and position-dependent magnetization, mass, and electromagnetic potential.

One finds
\begin{equation}
|| u^n||^2 + ||v^n||^2 \leq \frac{E_o}{1-\tilde{r}}, \mbox{for } \tilde{ r}=2 \sqrt{r_x^2+r_y^2}< 1 ~,\label{stab1}
\end{equation}
for scheme Eq. \eqref{lfsgscheme}, as well as 
\begin{equation}
|| \hat{u}^n||^2 + ||\hat{v}^n||^2 \leq \frac{\hat{E}_o}{1-\tilde{r}},  \mbox{for } \tilde{r}=2 \sqrt{r_x^2+r_y^2}< 1 ~,\label{stab2}
\end{equation}
for scheme Eqs. \eqref{uex1} and \eqref{vex1}, which is approximated by the scheme Eqs. \eqref{lfsgscheme-u} and  \eqref{lfsgscheme-v} above.
Stability can also be shown for $\tilde{r}=1$.
The proof is given in \ref{C}.

\begin{figure}[h!]
\centering
\includegraphics[width=8.5cm]{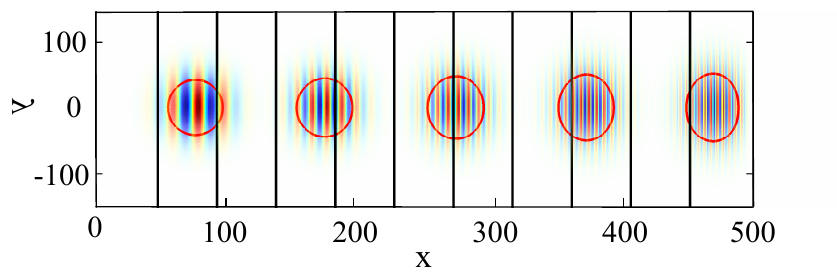}
\caption{(color online). Wave packet with $(k_x,k_y)=(10.0\pm1.3,0.0\pm1.3)\%$ for $m=0$ in a potential, shown by contour lines,  rising linearly in $x$-direction from $0$ to $0.2$. The closed lines represent the FWHM, the brightness saturation the probability density, and the color/brightness variation encodes the phase.}
\label{sim_constEfield}
\end{figure}
\section{Numerical examples}\label{numex}

With the numerical approach presented above, one is in a perfect position to simulate  ballistic Dirac fermion dynamics in (2+1)D in complex potential landscapes under open boundary conditions.  The numerical examples below have been selected mainly to demonstrate the properties of the scheme and provide some intuition for  (2+1)D ballistic Dirac fermion dynamics in simple effective TI mass and potential textures. 
For the examples given below, we have chosen a spatial region of $500\times500$ units and use a rather coarse discretization  $\Delta x = \Delta y = 1$, and $r=0.9$.  We give $k$ in percents ($\%$) of $k_{max}=\pi/\Delta$.   While movies are ideally suited to visualize our numerical results of Dirac fermion dynamics,   here the  figures show snapshots taken every $100$ time steps which are combined in a single plot.  The closed lines  in these figures mark the FWHM of the wave packet for  a given time step and the brightness saturation shows the probability density $\left\|\mbox{\boldmath$\psi$}\right\|^2$.  Color/brightness indicates the phase of the wave packet which is associated with the real part of the upper spinor component $u$.\\
\\

\subsection{Dispersive properties of the scheme}

Due to the chosen staggered-grid discretization a rather faithful representation of the exact continuum Dirac cone dispersion and constant "magnetization" ${\bf m}$,  
\begin{equation}
\Omega(k_x,k_y) =  \sqrt{\big(k_x-m_x\big)^2+\big(k_y-m_y)^2 + m_z^2} \;+ V\label{omegaconti}~,
\end{equation}
\\
is possible over a  wide region of k-space.
Eq. \eqref{omegaconti} is to be compared with the dispersion 
provided by the staggered grid discretization given in Eq. \eqref{grid-dis}.
\begin{figure}[t!]
\centering
\includegraphics[width=8.5cm]{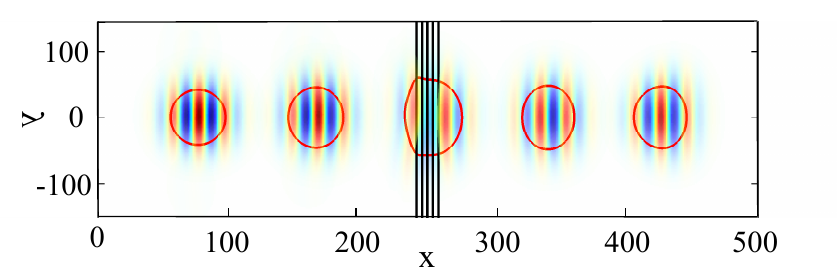}
\caption{(color online). Wave packet with $(k_x,k_y)=(10.0\pm1.3,0.0\pm1.3)\%$ for $m=0$ at a Klein step: on the left side the potential $V=0$   and rises linearly to $V=0.1$ at the right side of the figure, as shown by contour lines. The closed lines represent the FWHM, the brightness saturation gives the probability density, and the color/brightness variation encodes the phase.}
\label{sim_klein}
\end{figure}
\begin{figure}[!ht]
\centering
\includegraphics[width=8.5cm]{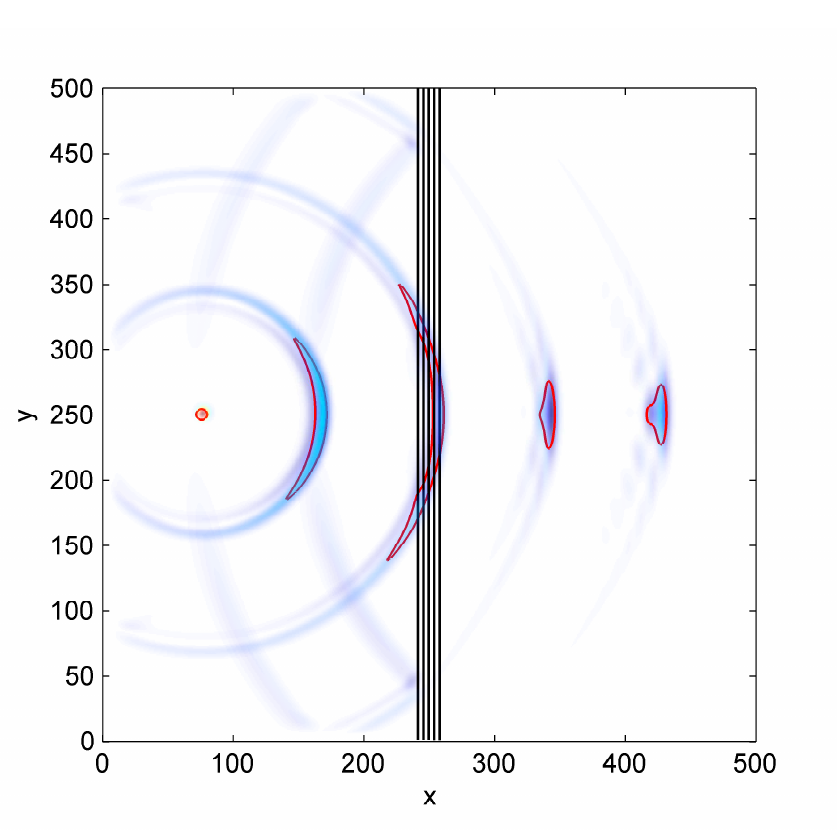}
\caption{(color online). Wave packet with $(k_x,k_y)=(5.0\pm6.4,0.0\pm6.4)\%$ for $m=0$ at a Klein step where on the left side $V=0$ and rising linearly to $V=0.1$ at the right side of the figure as shown by contour lines. The closed lines represent the FWHM, the brightness saturation gives the probability density, and the color/brightness variation encodes the phase.}
\label{sim_kleinlens}
\end{figure}
\begin{figure}[!ht]
\centering
\includegraphics[width=8.5cm]{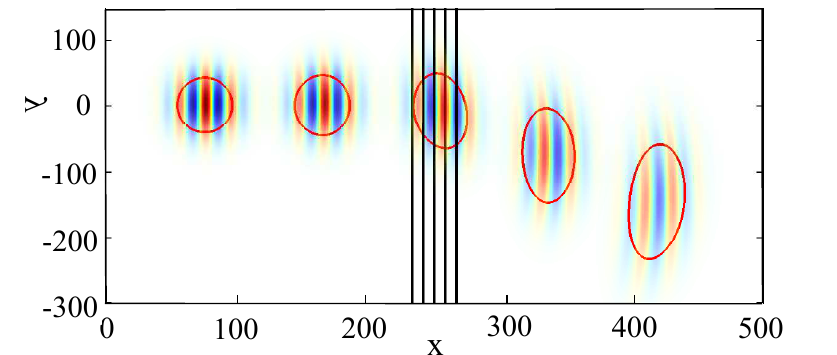}
\caption{(color online). Wave packet with $(k_x,k_y)=(10.0\pm1.3,0.0\pm1.3)\%$ for $V=0$ at a  step where on the left side $m_y=0$ and is rising linearly to $m_y=-0.2$ to the right side of the figure as shown by contour lines. The closed lines represent the FWHM, the brightness saturation gives the probability density, and the color/brightness variation encodes the phase.}
\label{sim_my}
\end{figure}
\begin{figure*}[t!]
\includegraphics[width=18cm]{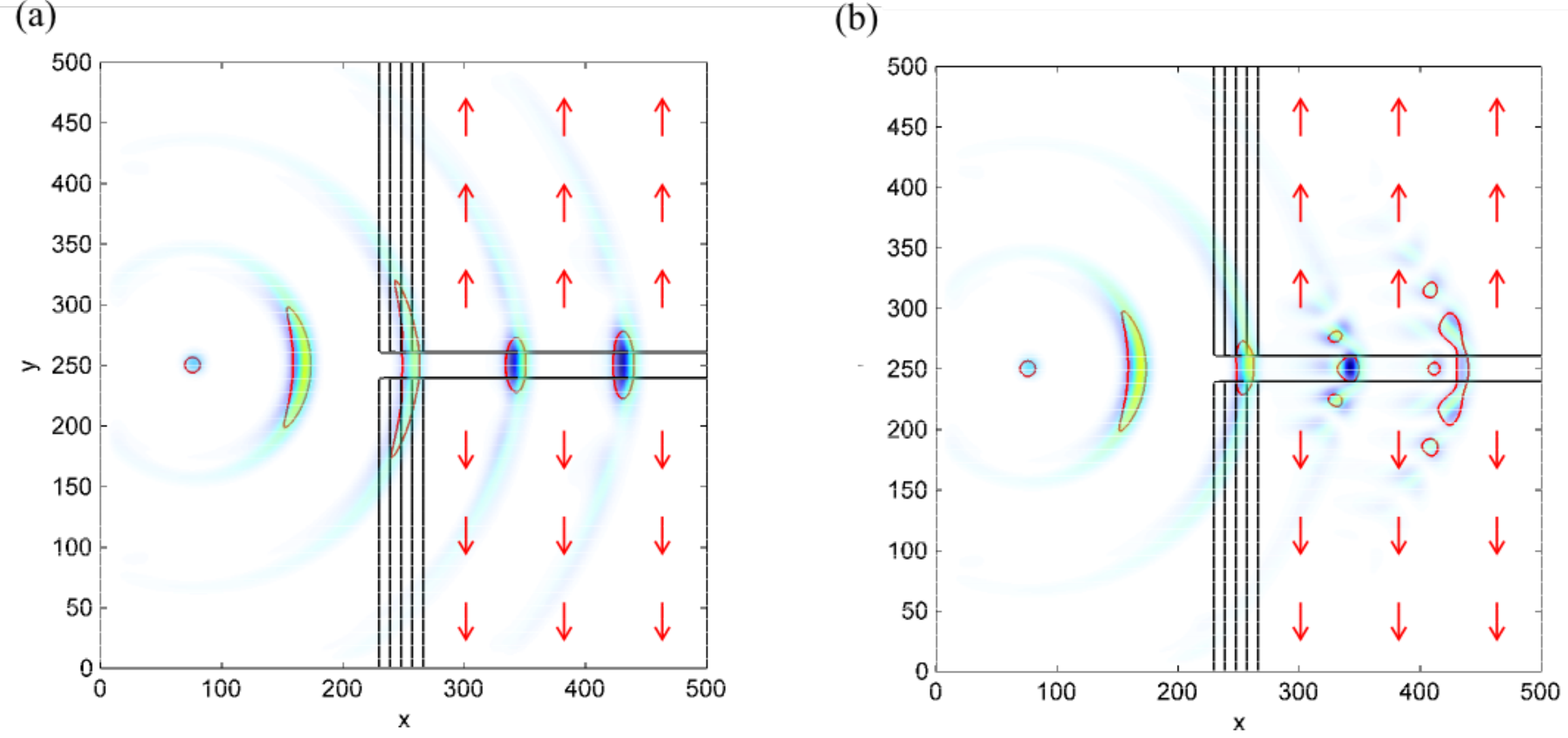}
\caption{(color online). (a) Wave packet with $(k_x,k_y)=(5.0\pm6.4,0.0\pm6.4)\%$ for $m=0$ at a  step where on the left side $m_y=0$ and rising linearly to $m_y\pm0.01$ (magnitude shown by contour lines and direction by arrows) where at the upper right-hand quarter of the figure $m_y=0.01$ and at the lower right-hand quarter of the figure $m_y=-0.01$. (b) the same as in (a) but with $m_y=0.03$. The closed lines represent the FWHM, the brightness saturation gives the probability density, and the color/brightness variation encodes the phase.}
\label{sim_mylenses}
\end{figure*}

\noindent First we demonstrate the quality of the dispersion on the lattice and investigate a "race" between two massless Dirac fermions on the grid. One is described by a wave packet with small wave numbers, prepared with a mean wave number of $(k_x,k_y)=(10.0,0.0)\%$ of $k_{max}$,  the other  by a wave packet with a high central wave number of  $k_{max}$: $(k_x,k_y)=(80.0, 0.0)\%$.  Both wave packets have a Gaussian half width in $k_x$ and $k_y$ of $(1.3,1.3)\%$ of $k_{max}$. Remember $k_{max}$ is the maximum wave number provided by the grid $k_{max}=\pi/\Delta$. Below we use the abbreviated notation $(k_x,k_y)=(80.0\pm1.3,0.0\pm1.3)\%$. Results are shown in Fig. \ref{sim_racecoarsefine} (a). Since the systematic errors in group velocity associated with the present scheme are small the two wave packets  propagate essentially with  equal speed. For this particular simulation we chose $r=1$ because it provides the best approximation  to the exact linear dispersion. The stronger distortion of the wave packet with the smaller wave number compared to the one with higher wave number is due to a different slope in the $y$-direction of the dispersion for the wave packet lying closer/farther to the center of the Dirac cone. This effect is also present for the continuum problem. To demonstrate this fact we show this "race" using the a very fine grid Fig. \ref{sim_racecoarsefine} (b).
\\
\\
\noindent In Fig. \ref{sim_constEfield} we show a wave packet of mass zero in a region of constant electric field represented by a linear  potential $V$ growing from $0$ to $0.1$. The initial Gaussian wave packet is prepared with $(k_x,k_y)=(10.0,0.0)\%\;k_{max}$ and a  half width of $(1.3,1.3)\%$. The speed of propagation does not change because the massless fermion on the grid always moves at maximum wave velocity ($c=1$). The change in kinetic energy shows up in the growth of the wave number. Here it is remarkable that the wave number grows close to the maximum wave number provided by the grid and yet the simulated propagation is still a good approximation to the exact result for the continuum problem. This is due to the ecellent dispersion properties of the scheme for wave numbers aligned with the grid.\\
\\

\subsection{Dynamics at Klein steps} 
\noindent Next we consider a Klein potential step at which the wave packet $(k_x,k_y)=(10.0\pm1.3,0.0\pm1.3)\%$ propagates in positive energy states on the left-hand side, where $V=0$, and in negative energy states on the right-hand side of the step, where $V=0.1$. 
For this simulation shown in Fig. \ref{sim_klein} we choose ${\bf m}=0$.  This situation leads to a high transmission of $\approx1$ of the wave packet under normal incidence since the potential step resonantly connects particle states on one side to hole (anti-particle) states on the other.  For the case $m_z\neq0$, as is well known, the transmission probability  grows with the height of the potential step \cite{itzykson}.\\
\\
\noindent  Figure Fig. \ref{sim_kleinlens}  shows the same Klein potential step but for an initial  wave packet $(k_x,k_y)=(5.0\pm6.4,0.0\pm6.4)\%$.  The observed  focusing behavior is a consequence of  phase and group velocity changing their sign across the step.
Also demonstrated in this figure is the successful implementation of absorbing boundary conditions.  Note that wave contributions impinging upon the simulation boundaries disappear without artificial reflection.

\subsection{Dynamics under finite ${\bf m}$} 

\noindent In this second part we explore simple cases for which ${\bf m}\neq 0$ emulating, for example, TI surfaces with ferromagnetic texture.  First we consider a situation  where $m_y(x)$ changes from zero to a finite constant value.  As seen in Eq. \eqref{omegaconti} a constant $m_y$ shifts the dispersion relation in $y$-direction of  momentum space.   Thus,  when a wave packet impinges upon such a "magnetic Klein step"  a wave packet starting with group velocity component $v_y=0$  ends up with a finite group velocity component, as demonstrated numerically in Fig. \ref{sim_my}.  
\noindent This effect can be exploited for the focusing of an incoming fermion beam, as shown in  Fig. \ref{sim_mylenses}.  We set, for the right upper quarter of the simulated domain,  $m_y=const$ and  $m_y=-const$ in the lower right quarter.    Therefore, for sufficiently small $k_y$,  the group velocity component  $v_y$ changes sign at the interface between these two regions. As a result, one observes the emergence of interference fringes where the distance of the maxima depends on $m_y$, see  Fig. \ref{sim_mylenses} (a) and (b).

\begin{figure}[!t]
\centering
\includegraphics[width=8.5cm]{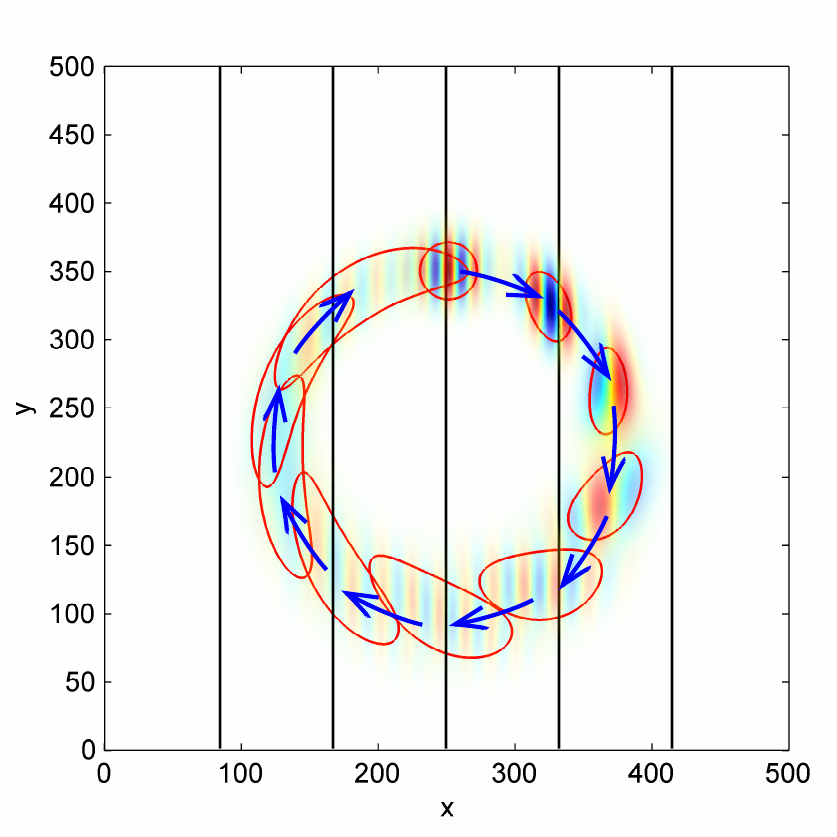}
\caption{(color online). Wave packet starting with $(k_x,k_y)=(10.0\pm0.8,0.0\pm0.8)\%$ for $V=0$ moving in a constant magnetic field in $z$-direction, represented by a  vector potential $m_y$  rising linearly with $x$. The closed lines mark the FWHM, the brightness saturation gives the probability density,  and the color/brightness variation encodes the phase.}
\label{sim_landau}
\end{figure}

\noindent Fig. \ref{sim_landau} shows a mass-zero wave packet traveling on a relativistic Landau orbit enforced by a magnetization in $z$-direction represented by a vector potential $m_y$ with linear $x$-dependence.  Positions of constant $m_y$  are indicated in the figure by vertical lines.  The Gaussian wave packet sets out in the 12 o'clock position and propagates in the direction indicated by the arrows.    In this case a strongly dispersive behavior is observed: inner portions of the wave packet move ahead of the outer portion leading to a significantly 
elongated wave packet after one completed orbit.  This dispersive effect is not a numerical error but a represents a peculiarity of a  mass zero particle in a magnetic field, for which the 
period $T$ for completion of one orbit fulfills  $T\propto k/B$, where $k$ is the magnitude of the k-vector and B the z-component of the magnetic field.  Since the particle moves at constant group velocity, also the 
classical cyclotron radius $R_c$ scales linearly with $k$.  Thus portions of  the wave packet closer to the center of the orbit move ahead of those further away from it, as is clearly revealed in the numerical simulation shown in Fig.  \ref{sim_landau}.   This effect is of course closely related to the $\sqrt{n}$ dependence of the cyclotron mass of graphene, with $n$ denoting the particle density, which has lead to the measurement of the Fermi velocity in graphene, as discussed in a recent review article \cite{neto}.  This effect can also be seen in a publication where a FFT-split-operator code was used \cite{mocken2}.

This numerical approach has also served as the basis for a theoretical study of Dirac fermion propagation on magnetically textured surfaces of topological insulators, in the vicinity of domain walls, domain wall intersections, and Dirac fermion wave guides \cite{hammerArXiV,hammerAPL}.

\section{Discussion}\label{discu} 

The main advantages from this compact scheme are the preservation of the free Weyl-Dirac fermion dispersion relation for wave vectors aligned along both the $x$- and the $y$-axis, as well as a large, nearly isotropic monotonic region near the center of k-space.   In particular for long-time simulations of magnetic textures  which are aligned along the main axes, like rectangular waveguide structures which we have modeled \cite{hammerAPL}, the advantages of the present scheme pay off. As demonstrated above, for such setups errors in phase and group velocity relative to the continuum model are small  (see, for example, the dispersion relation along the axes in Fig. \ref{disp-lf} (a)). The results show high accuracy for wave-components up to the grid maximum of two grid-points per wavelength, known as the Nyquist wave number.  It should also be remembered 
that the effective model captured by the Hamiltonian in Eq. \eqref{DHAM} is valid only in the vicinity of the degeneracy point of the 
surface states of a TI or the graphene band structure and for weak contributions from the ``magnetization" term.    Therefore, the presence of a second cone at $(k_x,k_y)=(\pm \frac{\pi}{\Delta x},\pm\frac{\pi}{\Delta y})$ does not provide a serious drawback for most applications.  

Furthermore, this scheme leaves open the option to choose, within its explicitly derived convergence limits,  any desirable ratios in the grid spacings. 
For simulations with arbitrary propagation directions one has to estimate the occurring wave numbers and choose the grid size such that the dispersion relation on the grid is a good approximation to the one for the continuum equation.  Again, the present scheme has a significant advantage (in addition of being explicit) over the standard scheme using a symmetric form of the spatial derivatives, see Fig. \ref{disp-lf} (b), since the region of k-space where a good approximation of the continuum energy dispersion is provided is significantly enlarged.  Typical simulations shown here take a few minutes of CPU time on an average PC. For this second order accurate scheme the numerical cost increases linearly with the number of (space+time) grid-points, offering a profound advantage compared to implicit schemes of the same order, whenever simulations of high accuracy are required.
If necessary, the second Dirac cone can be removed within this scheme by adding mass to the doublers using a Wilson term \cite{wilson}.  When properly implemented into the present staggered grid scheme it leads to an implicit scheme which will be shown elsewhere.

\section{Summary, conclusions, and outlook}\label{summa}
 
 In summary, we have presented a staggered-grid leap-frog scheme for the numerical solution of the (2+1)D Dirac equation which has the following favorable properties: it is an explicit scheme, it has the minimum number "two" of Dirac cones  on the lattice whereby the second cone sits at the corners of the ``1st Brillouin zone", for the case of mass zero it provides the correct linear dispersion along x- and y-direction, and it allows for the implementation of absorbing boundary conditions for simulations on a finite grid without spurious reflections from its boundaries,  
 as well as the simulation of particle sources and sinks.  Here, generic numerical examples have been given to demonstrate and explore these properties.  As a consequence, this approach is  well suited for the study of Dirac fermion dynamics in potential landscapes provided by external electromagnetic potentials, typical to TI surface states.
Applications to TI surface state dynamics based on this algorithm have and will be presented elsewhere \cite{hammerArXiV,hammerAPL}.   A related numerical treatment  of the (1+1)D two-spinor-component Dirac equation including {\it exact} absorbing boundary conditions,  displaying a single Dirac cone,  has been presented by us recently \cite{hammer1}.   Furthermore, we have been able to develop a scheme, respectively, for the to the (2+1)D two-spinor-component Dirac equation and the  (3+1)D four-spinor-component Dirac equation with a single cone only \cite{hammer2}.

\section*{Acknowledgments}
We acknowledge support from the Austrian Science Foundation under project I395-N16.\\  Furthermore, we thank B. A. Stickler, A. Arnold and C. Ertler for helpful comments.  


%
%
%
%
%
%
%
\newpage
\begin{onecolumn}
\begin{appendix}
\section{Scheme with Peierls substitution for the introduction of non-vanishing in-plane ``magnetization" $m_x$ and $m_y$\label{A}}

\noindent Here we show the consequence of the Peierls substitution Eq. \eqref{peierlss} into the scheme Eq. \eqref{lfsgscheme} to introduce non-vanishing 
``magnetization" $m_x$ and $m_y$ terms into the numerical scheme.  

We first explore the effect of the substitution on difference quotients (derivative terms)
\begin{equation}
\frac{u_1-u_2}{\Delta} \rightarrow \frac{e^{ia_1}u_1-e^{ia_2}u_2}{\Delta} ~.
\end{equation}
Simple regrouping gives the exact ``product rule for differentiation on the lattice"
\begin{equation}
\frac{e^{ia_1}u_1-e^{ia_2}u_2}{\Delta} =f_+(a_1,a_2)\frac{u_1-u_2}{\Delta}+\frac{e^{ia_1}-e^{ia_2}}{\Delta}\frac{u_1+u_2}{2}~,
\end{equation}
using the definition  $f_\pm(a_1,a_2)=(e^{ia_1}\pm e^{ia_2})/2$.
The last term on the rhs contains a difference quotient representing the derivative of an exponential.
It may be approximated by  the ``chain rule" for the derivative of exponentials on the grid 
\begin{equation}
\frac{e^{ia_1}-e^{ia_2}}{\Delta}= f_+(a_1,a_2)\frac{i(a_1-a_2)}{\Delta} + \frac{1}{\Delta}O((a_1-a_2)^3)~. 
\label{chain}
\end{equation}
The second type of terms to be dealt with are spatial averages of the structure
\begin{equation}
\frac{u_1+u_2}{2} \rightarrow \frac{e^{ia_1}u_1+e^{ia_2}u_2}{2}~.
\end{equation}
Here one arrives at
\begin{equation}
\frac{e^{ia_1}u_1+e^{ia_2}u_2}{2}=  f_+(a_1,a_2)\frac{u_1+u_2}{2} + f_-(a_1,a_2)\frac{u_1-u_2}{2}~.
\end{equation}\\
\\
The scheme which then arises from the Peierls substitution \eqref{peierlss} into Eq.  \eqref{lfsgscheme} is
\begin{align}
&f_+(a^{n+1}_{j,k},a^{n}_{j,k}) \left[\frac{u_{j,k}^{n+1}-u_{j,k}^{n}}{\Delta t}+ i\left((m_z)^{n}_{j,k}+V^{n}_{j,k} \right)\frac{u_{j,k}^{n+1}+u_{j,k}^{n}}{2} \right] \nonumber \\
&+\frac{e^{ia^{n+1}_{j,k}}-e^{ia^{n}_{j,k}}}{\Delta t} \frac{u_{j,k}^{n+1}+u_{j,k}^{n}}{2} 
+ if_-(a^{n+1}_{j,k},a^{n}_{j,k})\left((m_z)^{n}_{j,k}-V^{n}_{j,k} \right)\frac{u_{j,k}^{n+1}-u_{j,k}^{n}}{2} \nonumber \\
&+\frac{1}{2} f_+(a^{n}_{j,k-1},a^{n}_{j-1,k-1}) \frac{v_{j,k-1}^{n}-v_{j-1,k-1}^{n}}{\Delta x} + \frac{e^{ia^{n}_{j,k-1}}-e^{ia^{n}_{j-1,k-1}}}{ \Delta x}\frac{v_{j,k-1}^{n}+v_{j-1,k-1}^{n}}{4} 
   \nonumber \\
 &+\frac{1}{2} f_+(a^{n}_{j,k},a^{n}_{j-1,k})  \frac{v_{j,k}^{n}-v_{j-1,k}^{n}}{\Delta x} + \frac{e^{ia^{n}_{j,k}}-e^{ia^{n}_{j-1,k}}}{ \Delta x} \frac{v_{j,k}^{n}+v_{j-1,k}^{n}}{4} 
     \nonumber \\
 &- \frac{i}{2} f_+(a^{n}_{j-1,k},a^{n}_{j-1,k-1})  \frac{v_{j-1,k}^{n}-v_{j-1,k-1}^{n}}{\Delta y} -i \frac{e^{ia^{n}_{j-1,k}}-e^{ia^{n}_{j-1,k-1}}}{ \Delta y}\frac{v_{j-1,k}^{n}+v_{j-1,k-1}^{n}}{4} 
 \nonumber \\
&- \frac{i}{2} f_+(a^{n}_{j,k},a^{n}_{j,k-1}) \frac{v_{j,k}^{n}-v_{j,k-1}^{n}}{\Delta y} -i \frac{e^{ia^{n}_{j,k}}-e^{ia^{n}_{j,k-1}}}{ \Delta y}\frac{v_{j,k}^{n}+v_{j,k-1}^{n}}{4} 
  = 0~,\label{uex1}\qquad\qquad\qquad\qquad
\end{align}
and 
\begin{align}
&f_+(a^{n+1}_{j,k},a^{n}_{j,k}) \left[\frac{v_{j,k}^{n+1}-v_{j,k}^{n}}{\Delta t}- i\left((m_z)^{n+1}_{j,k}-V^{n+1}_{j,k} \right)\frac{v_{j,k}^{n+1}+v_{j,k}^{n}}{2} \right]\nonumber \\ 
&+\frac{e^{ia^{n+1}_{j,k}}-e^{ia^{n}_{j,k}}}{\Delta t}\frac{v_{j,k}^{n+1}+v_{j,k}^{n}}{2}
 - if_-(a^{n+1}_{j,k},a^{n}_{j,k})\left((m_z)^{n+1}_{j,k}+V^{n+1}_{j,k} \right)\frac{v_{j,k}^{n+1}-v_{j,k}^{n}}{2} \nonumber \\
&+\frac{1}{2} f_+(a^{n+1}_{j+1j,k},a^{n+1}_{j,k})  \frac{u_{j+1,k}^{n+1}-u_{j,k}^{n+1}}{\Delta x} +  \frac{e^{ia^{n+1}_{j+1,k}}-e^{ia^{n+1}_{j,k}}} {\Delta x} \frac{u_{j+1,k}^{n+1}+u_{j,k}^{n+1}}{4} 
  \nonumber \\
 &+\frac{1}{2} f_+(a^{n+1}_{j+1,k+1},a^{n+1}_{j,k+1}) \frac{u_{j+1,k+1}^{n+1}-u_{j,k+1}^{n+1}}{\Delta x} +  \frac{e^{ia^{n+1}_{j+1,k+1}}-e^{ia^{n+1}_{j,k+1}}} {\Delta x}  \frac{u_{j+1,k+1}^{n+1}+u_{j,k+1}^{n+1}}{4} 
   \nonumber \\
 &+ \frac{i}{2} f_+(a^{n+1}_{j,k+1},a^{n+1}_{j,k}) \frac{u_{j,k+1}^{n+1}-u_{j,k}^{n+1}}{\Delta y} +i  \frac{e^{ia^{n+1}_{j,k+1}}-e^{ia^{n+1}_{j,k}}} {\Delta y} \frac{u_{j,k+1}^{n+1}+u_{j,k}^{n+1}}{4} \nonumber \\
&+ \frac{i}{2} f_+(a^{n+1}_{j+1,k+1},a^{n+1}_{j+1,k}) \frac{u_{j+1,k+1}^{n+1}-u_{j+1,k}^{n+1}}{\Delta y} +i  \frac{e^{ia^{n+1}_{j+1,k+1}}-e^{ia^{n+1}_{j+1,k}}} {\Delta y}\frac{u_{j+1,k+1}^{n+1}+u_{j+1,k}^{n+1}}{4} 
  = 0~.  \label{vex1} 
\end{align}
It immediately becomes more transparent when the chain rule approximation Eq. \eqref{chain} is used
\begin{align}
&f_+(a^{n+1}_{j,k},a^{n}_{j,k}) \left[\frac{u_{j,k}^{n+1}-u_{j,k}^{n}}{\Delta t}+ i\left((m_z)^{n}_{j,k}+V^{n}_{j,k} +\frac{a^{n+1}_{j,k}-a^{n}_{j,k}}{\Delta t}\right)\frac{u_{j,k}^{n+1}+u_{j,k}^{n}}{2} \right] \nonumber \\
& + if_-(a^{n+1}_{j,k},a^{n}_{j,k})\left((m_z)^{n}_{j,k}-V^{n}_{j,k} \right)\frac{u_{j,k}^{n+1}-u_{j,k}^{n}}{2} \nonumber \\
&+\frac{1}{2} f_+(a^{n}_{j,k-1},a^{n}_{j-1,k-1}) \left[ \frac{v_{j,k-1}^{n}-v_{j-1,k-1}^{n}}{\Delta x} + \frac{v_{j,k-1}^{n}+v_{j-1,k-1}^{n}}{2} 
 \frac{i(a^{n}_{j,k-1}-a^{n}_{j-1,k-1})} {\Delta x}     \right ] \nonumber \\
 &+\frac{1}{2} f_+(a^{n}_{j,k},a^{n}_{j-1,k}) \left[ \frac{v_{j,k}^{n}-v_{j-1,k}^{n}}{\Delta x} + \frac{v_{j,k}^{n}+v_{j-1,k}^{n}}{2} 
 \frac{i(a^{n}_{j,k}-a^{n}_{j-1,k})} {\Delta x}     \right ] \nonumber \\
 &- \frac{i}{2} f_+(a^{n}_{j-1,k},a^{n}_{j-1,k-1}) \left[ \frac{v_{j-1,k}^{n}-v_{j-1,k-1}^{n}}{\Delta y} + \frac{v_{j-1,k}^{n}+v_{j-1,k-1}^{n}}{2} 
 \frac{i(a^{n}_{j-1,k}-a^{n}_{j-1,k-1})} {\Delta y} \right]  \nonumber \\
&- \frac{i}{2} f_+(a^{n}_{j,k},a^{n}_{j,k-1}) \left[ \frac{v_{j,k}^{n}-v_{j,k-1}^{n}}{\Delta y} + \frac{v_{j,k}^{n}+v_{j,k-1}^{n}}{2} 
 \frac{i(a^{n}_{j,k}-a^{n}_{j,k-1})} {\Delta y} \right] = 0~,\qquad\qquad\qquad\qquad\quad
\end{align}
and
\begin{align}
&f_+(a^{n+1}_{j,k},a^{n}_{j,k}) \left[\frac{v_{j,k}^{n+1}-v_{j,k}^{n}}{\Delta t}- i\left((m_z)^{n+1}_{j,k}-V^{n+1}_{j,k} -\frac{a^{n+1}_{j,k}-a^{n}_{j,k}}{\Delta t}\right)\frac{v_{j,k}^{n+1}+v_{j,k}^{n}}{2} \right]\nonumber \\ 
& - if_-(a^{n+1}_{j,k},a^{n}_{j,k})\left((m_z)^{n+1}_{j,k}+V^{n+1}_{j,k} \right)\frac{v_{j,k}^{n+1}-v_{j,k}^{n}}{2} \nonumber \\
&+\frac{1}{2} f_+(a^{n+1}_{j+1j,k},a^{n+1}_{j,k}) \left[ \frac{u_{j+1,k}^{n+1}-u_{j,k}^{n+1}}{\Delta x} + \frac{u_{j+1,k}^{n+1}+u_{j,k}^{n+1}}{2} 
 \frac{i(a^{n+1}_{j+1,k}-a^{n+1}_{j,k})} {\Delta x}     \right ] \nonumber \\
 &+\frac{1}{2} f_+(a^{n+1}_{j+1,k+1},a^{n+1}_{j,k+1}) \left[ \frac{u_{j+1,k+1}^{n+1}-u_{j,k+1}^{n+1}}{\Delta x} + \frac{u_{j+1,k+1}^{n+1}+u_{j,k+1}^{n+1}}{2} 
 \frac{i(a^{n+1}_{j+1,k+1}-a^{n+1}_{j,k+1})} {\Delta x}     \right ] \nonumber \\
 &+ \frac{i}{2} f_+(a^{n+1}_{j,k+1},a^{n+1}_{j,k}) \left[ \frac{u_{j,k+1}^{n+1}-u_{j,k}^{n+1}}{\Delta y} + \frac{u_{j,k+1}^{n+1}+u_{j,k}^{n+1}}{2} 
 \frac{i(a^{n+1}_{j,k+1}-a^{n+1}_{j,k})} {\Delta y} \right]  \nonumber \\
&+ \frac{i}{2} f_+(a^{n+1}_{j+1,k+1},a^{n+1}_{j+1,k}) \left[ \frac{u_{j+1,k+1}^{n+1}-u_{j+1,k}^{n+1}}{\Delta y} + \frac{u_{j+1,k+1}^{n+1}+u_{j+1,k}^{n+1}}{2} 
 \frac{i(a^{n+1}_{j+1,k+1}-a^{n+1}_{j+1,k})} {\Delta y} \right] = 0~.
\end{align}

\noindent Consistent with the ``chain rule", under weak spatial and temporal variation of the magnetization (vector potential) ${\bf m}(x,y,t)$,  the $f_-$  terms may be dropped and the  $f_+$ factors may be eliminated, leading to 
a simplified version of the form
\begin{align}
&\frac{u_{j,k}^{n+1}-u_{j,k}^{n}}{\Delta t}+ i\left((m_z)^{n}_{j,k}+V^{n}_{j,k} +\frac{a^{n+1}_{j,k}-a^{n}_{j,k}}{\Delta t}\right)\frac{u_{j,k}^{n+1}+u_{j,k}^{n}}{2}  \nonumber \\
&+\frac{1}{2} \left[ \frac{v_{j,k-1}^{n}-v_{j-1,k-1}^{n}}{\Delta x} + \frac{v_{j,k-1}^{n}+v_{j-1,k-1}^{n}}{2} 
 \frac{i(a^{n}_{j,k-1}-a^{n}_{j-1,k-1})} {\Delta x}     \right ] \nonumber \\
 &+\frac{1}{2} \left[ \frac{v_{j,k}^{n}-v_{j-1,k}^{n}}{\Delta x} + \frac{v_{j,k}^{n}+v_{j-1,k}^{n}}{2} 
 \frac{i(a^{n}_{j,k}-a^{n}_{j-1,k})} {\Delta x}     \right ] \nonumber \\
 &- \frac{i}{2}  \left[ \frac{v_{j-1,k}^{n}-v_{j-1,k-1}^{n}}{\Delta y} + \frac{v_{j-1,k}^{n}+v_{j-1,k-1}^{n}}{2} 
 \frac{i(a^{n}_{j-1,k}-a^{n}_{j-1,k-1})} {\Delta y} \right]  \nonumber \\
&- \frac{i}{2} \left[ \frac{v_{j,k}^{n}-v_{j,k-1}^{n}}{\Delta y} + \frac{v_{j,k}^{n}+v_{j,k-1}^{n}}{2} 
 \frac{i(a^{n}_{j,k}-a^{n}_{j,k-1})} {\Delta y} \right] = 0~,\qquad\qquad\qquad
\end{align}
\begin{align}
&\frac{v_{j,k}^{n+1}-v_{j,k}^{n}}{\Delta t}- i\left((m_z)^{n+1}_{j,k}-V^{n+1}_{j,k} -\frac{a^{n+1}_{j,k}-a^{n}_{j,k}}{\Delta t}\right)\frac{v_{j,k}^{n+1}+v_{j,k}^{n}}{2}  \nonumber \\
&+\frac{1}{2}  \left[ \frac{u_{j+1,k}^{n+1}-u_{j,k}^{n+1}}{\Delta x} + \frac{u_{j+1,k}^{n+1}+u_{j,k}^{n+1}}{2} 
 \frac{i(a^{n+1}_{j+1,k}-a^{n+1}_{j,k})} {\Delta x}     \right ] \nonumber \\
 &+\frac{1}{2} \left[ \frac{u_{j+1,k+1}^{n+1}-u_{j,k+1}^{n+1}}{\Delta x} + \frac{u_{j+1,k+1}^{n+1}+u_{j,k+1}^{n+1}}{2} 
 \frac{i(a^{n+1}_{j+1,k+1}-a^{n+1}_{j,k+1})} {\Delta x}     \right ] \nonumber \\
 &+ \frac{i}{2}  \left[ \frac{u_{j,k+1}^{n+1}-u_{j,k}^{n+1}}{\Delta y} + \frac{u_{j,k+1}^{n+1}+u_{j,k}^{n+1}}{2} 
 \frac{i(a^{n+1}_{j,k+1}-a^{n+1}_{j,k})} {\Delta y} \right]  \nonumber \\
&+ \frac{i}{2}  \left[ \frac{u_{j+1,k+1}^{n+1}-u_{j+1,k}^{n+1}}{\Delta y} + \frac{u_{j+1,k+1}^{n+1}+u_{j+1,k}^{n+1}}{2} 
 \frac{i(a^{n+1}_{j+1,k+1}-a^{n+1}_{j+1,k})} {\Delta y} \right] = 0~,
\end{align}
Here  $\hat{V}^{n(+1)}_{j,k} = V^{n(+1)}_{j,k} +\frac{a^{n+1}_{j,k}-a^{n}_{j,k}}{\Delta t}$ denotes the net scalar potential in presence of a vector potential.  The 
$x$ and $y$ components of the magnetization (vector potential) on the grid, respectively, are given by 
\begin{align*}
(m_x)^n_{j,k}&\approx \frac{a^n_{j,k}-a^n_{j-1,k}}{\Delta x}\approx  \frac{a^n_{j,k-1}-a^n_{j-1,k-1}}{\Delta x}~,\\
(m_y)^{n}_{j,k}&\approx \frac{a^n_{j-1,k}-a^n_{j-1,k-1}}{\Delta y}\approx  \frac{a^n_{j,k}-a^n_{j-1,k-1}}{\Delta y}~,\\
(m_x)^{n+1}_{j,k}&\approx \frac{a^{n+1}_{j+1,k}-a^{n+1}_{j,k}}{\Delta x}\approx  \frac{a^{n+1}_{j+1,k+1}-a^{n+1}_{j,k+1}}{\Delta x}~,
\end{align*}
and
\begin{equation}
(m_y)^{n+1}_{j,k}\approx \frac{a^{n+1}_{j,k+1}-a^{n+1}_{j,k}}{\Delta y}\approx  \frac{a^{n+1}_{j+1,k+1}-a^{n+1}_{j+1,k}}{\Delta y}~.
\end{equation}

\noindent Using these simplifications we finally arrive at the scheme  Eqs. \eqref{lfsgscheme-u} and \eqref{lfsgscheme-v}
\begin{align}
\frac{u_{j,k}^{n+1}-u_{j,k}^{n}}{\Delta t}=&- i\left((m_z)^{n}_{j,k}+\hat{V}^{n}_{j,k} \right)\frac{u_{j,k}^{n+1}+u_{j,k}^{n}}{2} \nonumber \\
 &-\frac{v_{j,k-1}^{n}-v_{j-1,k-1}^{n}+v_{j,k}^{n}-v_{j-1,k}^{n}}{2 \Delta x} 
 + i \frac{v_{j-1,k}^{n}-v_{j-1,k-1}^{n}+v_{j,k}^{n}-v_{j,k-1}^{n}}{2\Delta y}\nonumber \\
 &- i(m_x)^{n}_{j,k}  \frac{v_{j,k}^{n}+v_{j-1,k}^{n}+ v_{j,k-1}^{n}+v_{j-1,k-1}^{n}}{4}
 -(m_y)^{n}_{j,k} \frac{v_{j-1,k}^{n}+v_{j-1,k-1}^{n}+v_{j,k}^{n}+v_{j,k-1}^{n}}{4}~,
\end{align}
and
\begin{align}
\frac{v_{j,k}^{n+1}-v_{j,k}^{n}}{\Delta t}= &+ i\left(m_z-\hat{V}_{j,k}^{n+1} \right)\frac{v_{j,k}^{n+1}+v_{j,k}^{n}}{2}  \nonumber \\
&- \frac{u_{j+1,k}^{n+1}-u_{j,k}^{n+1}+u_{j+1,k+1}^{n+1}-u_{j,k+1}^{n+1}}{2\Delta x}
-\frac{u_{j,k+1}^{n+1}-u_{j,k}^{n+1}+u_{j+1,k+1}^{n+1}-u_{j+1,k}^{n+1}}{2\Delta y} \nonumber \\
& -i(m_x)^{n+1}_{j,k}\frac{u_{j+1,k}^{n+1}+u_{j,k}^{n+1}+u_{j+1,k+1}^{n+1}+u_{j,k+1}^{n+1}}{4} 
+ (m_y)^{n+1}_{j,k}\frac{u_{j,k+1}^{n+1}+u_{j,k}^{n+1}+u_{j+1,k+1}^{n+1}+u_{j+1,k}^{n+1}}{4} ~.
\end{align}

\newpage
\section{Derivation of a functional for the norm which is exactly conserved by the scheme\label{B}}

\noindent The proof of Eq. \eqref{norm} can be given as follows \cite{hammer1}.  Form the scalar product of the  first of equation Eqs. \eqref{lfsgscheme}  with $(u^{n+1}+u^{n})$, with the latter applied from the left,  and retain the real part of the resulting equation. 
This gives, using the scalar products introduced above Eqs. \eqref{scp1} to \eqref{scp3},
\begin{align}
\frac{|| u^{n+1}||^2 - ||u^n||^2}{\Delta t}= &-\Re\left\{\frac{(u^{n+1}+u^n;v^n)_{0,-} - (u^{n+1}+u^n;v^n)_{-,-} + (u^{n+1}+u^n;v^n)_{0,0} - (u^{n+1}+u^n;v^n)_{-,0}}{2\Delta x}\right\} \nonumber \\
&-\Im\left\{\frac{(u^{n+1}+u^n;v^n)_{-,0} - (u^{n+1}+u^n;v^n)_{-,-} + (u^{n+1}+u^n;v^n)_{0,0}-(u^{n+1}+u^n;v^n)_{0,-}}{2\Delta y}\right\}~.
\end{align}
Form the scalar product between the second equation of Eqs. \eqref{lfsgscheme} with $( v^{n+1}+v^{n})$, with the latter applied from the right, and take the real part of this equation.  This gives
\begin{align}
\frac{|| v^{n+1}||^2 - ||v^n||^2}{\Delta t}= &-\Re\left\{\frac{(u^{n+1};v^{n+1}+v^n)_{-,0} - (u^{n+1};v^{n+1}+v^n)_{0,0} + (u^{n+1};v^{n+1}+v^n)_{-,-} - (u^{n+1};v^{n+1}+v^n)_{0,-}}{2\Delta x}\right\} \nonumber \\
&-\Im\left\{\frac{(u^{n+1};v^{n+1}+v^n)_{0,-} - (u^{n+1};v^{n+1}+v^n)_{0,0} + (u^{n+1};v^{n+1}+v^n)_{-,-}-(u^{n+1};v^{n+1}+v^n)_{-,0}}{2\Delta y}\right\}~.
\end{align}

\noindent Adding these two equations yields, on the lhs,  $|| u^{n+1}||^2 + ||v^{n+1}||^2 - || u^n||^2 + ||v^n||^2$.  On the rhs one finds three types of terms: scalar products between equal-time components (respectively $n$ with $n$ and $n+1$ with $n+1$) and those mixing $n$ with $n+1$.   The latter cancel in pairs using a shift of indices,  such as $( u^n; v^{n'})_{-,0}-( u^n; v^{n'})= \sum_{jk} ( u_{j,k}^n ;v_{j-1,k}^{n'} - v_{j,k}^{n'})= \sum_{jk} ( u_{j+1,k}^n -u_{j,k}^n; v_{j,k}^{n'})$.   The $m_z$ and scalar potential terms vanish when taking the real part.
One obtains,
\begin{align}
|| u^{n+1}||^2 - ||u^n||^2+|| v^{n+1}||^2 - ||v^n||^2 =   &-r_x\Re\left\{(u^n;v^n)_{0,-} - (u^n;v^n)_{-,-} + 
(u^n;v^n)_{0,0} - (u^n;v^n)_{-,0}\right\} \nonumber \\
&-r_y\Im\left\{(u^n;v^n)_{-,0} - (u^n;v^n)_{-,-} + (u^n;v^n)_{0,0}-(u^n;v^n)_{0,-}\right\} \nonumber \\
&+r_x\Re\left\{(u^{n+1};v^{n+1})_{-,0} - (u^{n+1};v^{n+1})_{0,0} + (u^{n+1};v^{n+1})_{-,-} - (u^{n+1};v^{n+1})_{0,-}\right\} \nonumber \\
&+r_y\Im\left\{(u^{n+1};v^{n+1})_{0,-} - (u^{n+1};v^{n+1})_{0,0} + (u^{n+1};v^{n+1})_{-,-}-(u^{n+1};v^{n+1})_{-,0}\right\}~.
\end{align}
This is the identity $E_{n+1}$=$E_{n}$.  

In case of space and/or time dependent $m_x$ or $m_y$, an exactly conserved quantity is not obtained in this fashion from the  Eqs.    \eqref{lfsgscheme-u} and \eqref{lfsgscheme-v}.  However, it is readily constructed for the ``exact" equations obtained within the Peierls substitution, Eqs.  \eqref{uex1} and \eqref{vex1}, by simply  applying it to $E_n$ to give $\hat{E}_n$ which has the exact structure of $E_n$, just with $u,v$  replaced by $\hat{u},\hat{v}$.

\newpage
\section{Stability\label{C}}

\noindent In this appendix we prove the stability conditions for arbitrary space- and time-dependent magnetization vector and potential terms, Eq. \ref{stab1} and Eq. \ref{stab2}.
We use norm conservation Eq. \eqref{norm} and proceed as follows
\begin{align}
E_0 =E_n=|| u^n||^2 + ||v^n||^2 &- r_x\Re\left\{ (u^n;v^n)_{0,-} - (u^n;v^n)_{-,-} + (u^n;v^n)_{0,0} - (u^n;v^n)_{-,0} \right\} \nonumber  \\
&- r_y\Im\left\{ (u^n;v^n)_{-,0} - (u^n;v^n)_{-,-} + (u^n;v^n)_{0,0}-(u^n;v^n)_{0,-} \right\}\nonumber \\
\nonumber \\
=|| u^n||^2 + ||v^n||^2  &- \Re\left\{ (r_x+ir_y) \left[(u^n;v^n)_{0,-}- (u^n;v^n)_{-,0}\right]  - (r_x-ir_y)\left[(u^n;v^n)_{-,-} - (u^n;v^n)_{0,0} \right] \right\} \nonumber  \\
\nonumber \\
\geq || u^n||^2 + ||v^n||^2 &- \Big|\Re\left\{ (r_x+ir_y) \left[(u^n;v^n)_{0,-}- (u^n;v^n)_{-,0}\right]  - (r_x-ir_y)\left[(u^n;v^n)_{-,-} - (u^n;v^n)_{0,0} \right] \Big|  \right\} \nonumber  \\
\nonumber \\
\geq || u^n||^2 + ||v^n||^2 &- \mid \Re\left\{ (r_x+ir_y) (u^n;v^n)_{0,-}\right\} \mid -  \mid \Re\left\{(r_x+ir_y) (u^n;v^n)_{-,0}\right]\}\mid  \nonumber \\
&- \mid \Re\left\{ (r_x-ir_y)u^n;v^n)_{-,-}\right\}\mid - \mid \Re\left\{ (r_x-ir_y)(u^n;v^n)_{0,0} \right\} \mid \nonumber  \\
\nonumber \\
\geq || u^n||^2 + ||v^n||^2 &- \sqrt{r_x^2+r_y^2}\Big[\mid \Re\left\{ (u_-^n;v^n)_{0,-}\right\} \mid +  \mid \Re\left\{ (u_-^n;v^n)_{-,0}\right\}\mid \nonumber \\
&  \qquad\qquad\quad+ \mid \Re\left\{(u_+^n;v^n)_{-,-}\right\}\mid + \mid \Re\left\{ (u_+^n;v^n)_{0,0} \right\} \mid\Big] \nonumber  \\
\nonumber \\
\geq || u^n||^2 + ||v^n||^2 &- 2 \sqrt{r_x^2+r_y^2}\left( || u^n||^2 + ||v^n||^2 \right)=(1-\tilde{r})\left( || u^n||^2 + ||v^n||^2 \right) \nonumber. 
\end{align}
Here we have used the inequality $2\mid\Re\left\{(a,b)\right\}\mid \leq  || a ||^2 + ||b||^2$, as well as the abbreviation $u_\pm^n=e^{\pm i\phi}u_n$, where $\phi=\arctan{(r_y/r_x)}$.  Note also that the norm of a spinor component shifted by  $\pm \Delta x, \pm \Delta y$ is equal to the norm of the unshifted component under zero or periodic boundary conditions.  

The case $\tilde{r}=1$ can be dealt with as follows (omitting the superscript $n$ for brevity).  Starting from the last identity in the previous proof, for this case one may write
\begin{align}
E_0=E_n=|| u||^2 + ||v||^2  &+ \frac{1}{2} \Re\left\{  (u_-;v)_{-,0}-(u_-;v)_{0,-} + (u_+;v)_{-,-} - (u_+;v)_{0,0} \right\} \nonumber ~\\
&= \frac{1}{4} \left[ \mid\mid u_++v_{-,-}\mid\mid^2 +\mid\mid u_+-v\mid\mid^2 +\mid\mid u_-+v_{-,0}\mid\mid^2 +\mid\mid u_--v_{0,-}\mid\mid^2 \right]
\end{align}
Note once more, that the single subscript on the spinor component ($u$) indicates a phase shift, while a double subscript indicates a shift on the spatial grid (here applied to component $v$).  Here we use identities  of the form 
$$
\frac{1}{4}\left[||u_\pm + v||^2+ ||u_\pm - v'||^2\right] =\frac{1}{2} \left[||u||^2+ ||v||^2+ \Re\left\{(u_\pm,v)-(u_\pm,v')\right\}\right]
$$
when $ ||v||= ||v'||$.  Note that this identity can be  applied no more than twice, since $|| u||^2 + ||v||^2 $ is available in $E_n=E_0$.  This limits the magnitude of $\tilde{r}$.  In the second step we apply the inequality $2\left[ ||a_1+b||^2+||a_2-b||^2\right] \geq  ||a_1+a_2||^2$, whereby $u_\pm$ plays the role of $b$,  to obtain
$$
E_0\geq \frac{1}{8}\left[ || v_{-,-}+v||^2 + || v_{0,-}+v_{-,0}||^2 \right] \equiv ||\tilde{v}||^2~.
$$
Attaching the phase factor of $r_\pm$ to the components $v$ and the grid shifts to the components $u$, one obtains
$$
E_0\geq \frac{1}{8}\left[ || u_{+,+}+u|^2 + || u_{0,+}+u_{+,0}||^2 \right] \equiv ||\tilde{u}||^2.
$$
Hence
$$
||\tilde{u}||^2 +||\tilde{v}||^2 \leq 2E_0 ~.
$$
This discussion also shows that $E_n=E_0\geq 0$ for $\tilde{r}\leq 1$ and thus provides a meaningful definition for a spinor norm within the scheme.

With $E_0$, $u^n$, and $v^n$, respectively,  replaced by 
$\hat{E}_0$, $\hat{u}^n$, and $\hat{v}^n$ the stability condition for Eqs. \eqref{uex1} and \eqref{vex1} is shown under arbitrary space-time dependence of the external fields. 
\end{appendix}
\end{onecolumn}

\end{document}